\documentclass[a4paper,11pt]{article}
\pdfoutput=1 
\usepackage{jheppub}

\makeatletter
\def\@fpheader{}
\makeatother

\usepackage[T1]{fontenc}
\usepackage[utf8]{inputenc}
\usepackage{lmodern}

\usepackage{tikz} 
\usepackage{xcolor}
\usepackage{graphicx}
\graphicspath{{Figures/}}
\usepackage[english]{babel}
\usepackage{microtype}
\usepackage{amsmath,amssymb}
\usepackage{float}
\usepackage{bm}
\usepackage{enumitem}
\usepackage[hang,flushmargin]{footmisc}
\usepackage{relsize,exscale}
\usepackage{fancyhdr}
\usepackage{caption}
\usepackage{subcaption}
\usepackage{changepage}
\usepackage[most]{tcolorbox}
\tcbset{colback=yellow!10!white, colframe=red!50!black, 
	highlight math style= {enhanced, %<-- needed for the ’remember’ options
		colframe=red,colback=red!10!white,boxsep=0pt}}
\usepackage{empheq}
\usepackage{gensymb}
\usepackage{dsfont}
\usepackage[only,llbracket,rrbracket]{stmaryrd}
\usepackage{mathtools}
\usepackage{slashed}
\usepackage{hyperref}
\hypersetup{
	pagebackref=true,
	colorlinks=true,
	linktoc=section,   
	citecolor=blue, 
	filecolor=blue,
	linkcolor=blue,
	urlcolor=blue,
	pdfborder={0 0 0.5},
	pdfmenubar=false,
	pdftoolbar=false
}
\usepackage{cleveref}

\numberwithin{equation}{section}
\providecommand{\delimsize}{\relax}

\DeclarePairedDelimiterX{\bra}[1]{\delimsize\langle}{\delimsize\rvert}{#1}
\DeclarePairedDelimiterX{\ket}[1]{\delimsize\lvert}{\delimsize\rangle}{#1}
\DeclarePairedDelimiterX{\makebraket}[1]{\delimsize\langle}{\delimsize\rangle}{#1}

\makeatletter
\newcommand{\braketbar}{%
	\, \delimsize\vert\@ifnextchar|{\!}{\,\!\!\:}%
}
\makeatother
\newcommand{\activatebraketbar}{%
	\begingroup\lccode`~=`|\lowercase{\endgroup\let~}\braketbar
	\mathcode`|="8000
}

\NewDocumentCommand{\braket}{som}{%
	\mathord{%
		\begingroup
		\activatebraketbar
		\IfBooleanTF{#1}
		{\makebraket*{#3}}
		{\IfNoValueTF{#2}{\makebraket{#3}}{\makebraket[#2]{#3}}}%
		\endgroup
	}%
}
\newcommand{\ols}[1]{\mskip.5\thinmuskip\overline{\mskip-.5\thinmuskip {#1} \mskip-.5\thinmuskip}\mskip.5\thinmuskip} % overline short
\newcommand{\olsi}[1]{\,\overline{\!{#1}}} % overline short italic
\makeatletter
\newcommand\closure[1]{
	\tctestifnum{\count@stringtoks{#1}>1} %checks if number of chars in arg > 1 (including '\')
	{\ols{#1}} %if arg is longer than just one char, e.g. \mathbb{Q}, \mathbb{F},...
	{\olsi{#1}} %if arg is just one char, e.g. K, L,...
}
\long\def\count@stringtoks#1{\tc@earg\count@toks{\string#1}}
\long\def\count@toks#1{\the\numexpr-1\count@@toks#1.\tc@endcnt}
\long\def\count@@toks#1#2\tc@endcnt{+1\tc@ifempty{#2}{\relax}{\count@@toks#2\tc@endcnt}}
\def\tc@ifempty#1{\tc@testxifx{\expandafter\relax\detokenize{#1}\relax}}
\long\def\tc@earg#1#2{\expandafter#1\expandafter{#2}}
\long\def\tctestifnum#1{\tctestifcon{\ifnum#1\relax}}
\long\def\tctestifcon#1{#1\expandafter\tc@exfirst\else\expandafter\tc@exsecond\fi}
\long\def\tc@testxifx{\tc@earg\tctestifx}
\long\def\tctestifx#1{\tctestifcon{\ifx#1}}
\long\def\tc@exfirst#1#2{#1}
\long\def\tc@exsecond#1#2{#2}
\makeatother

\makeatletter
\newlength\xvec@height%
\newlength\xvec@depth%
\newlength\xvec@width%
\newcommand{\xvec}[2][]{%
	\ifmmode%
	\settoheight{\xvec@height}{$#2$}%
	\settodepth{\xvec@depth}{$#2$}%
	\settowidth{\xvec@width}{$#2$}%
	\else%
	\settoheight{\xvec@height}{#2}%
	\settodepth{\xvec@depth}{#2}%
	\settowidth{\xvec@width}{#2}%
	\fi%
	\def\xvec@arg{#1}%
	\def\xvec@dd{:}%
	\def\xvec@d{.}%
	\raisebox{.2ex}{\raisebox{\xvec@height}{\rlap{%
				\kern.05em%  (Because left edge of drawing is at .05em)
				\begin{tikzpicture}[scale=1]
					\pgfsetroundcap
					\draw (.05em,0)--(\xvec@width-.05em,0);
					\draw (\xvec@width-.05em,0)--(\xvec@width-.15em, .1em);
					\draw (\xvec@width-.05em,0)--(\xvec@width-.15em,-.1em);
					\ifx\xvec@arg\xvec@d%
					\fill(\xvec@width*.45,.5ex) circle (.5pt);%
					\else\ifx\xvec@arg\xvec@dd%
					\fill(\xvec@width*.30,.5ex) circle (.5pt);%
					\fill(\xvec@width*.65,.5ex) circle (.5pt);%
					\fi\fi%
				\end{tikzpicture}%
	}}}%
	#2%
}
\makeatother

\usepackage{contour}
\usepackage[normalem]{ulem}

\contourlength{0.8pt}

\makeatletter
\g@addto@macro\bfseries{\boldmath}
\makeatother

\newcounter{daggerfootnote}

\makeatletter
\DeclareRobustCommand{\cbig}[1]{\mathopen{\setbox0=\hbox{$\m@th\big#1$}\raisebox{-0.55pt}{$\scalebox{1.2}{\copy0}$}}}
\makeatother
\makeatletter
\DeclareRobustCommand{\cBig}[1]{\mathopen{\setbox0=\hbox{$\m@th\Big#1$}\raisebox{-0.4pt}{$\scalebox{1.15}{\copy0}$}}}
\makeatother
\makeatletter
\DeclareRobustCommand{\cbigg}[1]{\mathopen{\setbox0=\hbox{$\m@th\bigg#1$}\raisebox{-0.4pt}{$\scalebox{1.125}{\copy0}$}}}
\makeatother
\makeatletter
\DeclareRobustCommand{\cBigg}[1]{\mathopen{\setbox0=\hbox{$\m@th\Bigg#1$}\raisebox{-0.5pt}{$\scalebox{1.125}{\copy0}$}}}
\makeatother

\tikzset{
	cross/.pic = {
		\draw[rotate = 45] (-#1,0) -- (#1,0);
		\draw[rotate = 45] (0,-#1) -- (0, #1);
	}
}

\usetikzlibrary{decorations.pathmorphing}
\usetikzlibrary{arrows.meta,bending}
\title{Zero-damped modes of near-extremal Reissner--Nordstr\"{o}m black holes from exact WKB}

\author[a,b]{Prisco Lo Chiatto,}
\author[c]{Sebastian Schenk,}
\author[a,d]{Nils Wagner,}
\author[e]{and Felix Yu}
\affiliation[a]{Max Planck Institute for Physics (Werner Heisenberg Institute), Boltzmannstraße 8, 85748 Garching, Germany}
\affiliation[b]{Department of Particle Physics and Astrophysics, Weizmann Institute of Science, Rehovot 7610001, Israel}
\affiliation[c]{Institute for Astroparticle Physics, Karlsruhe Institute of Technology, Hermann-von-Helmholtz-Platz 1, 76344 Eggenstein-Leopoldshafen, Germany}
\affiliation[d]{Department of Physics, Technical University of Munich, James-Franck-Straße 1, 85748 Garching, Germany}
\affiliation[e]{PRISMA$^{++}$ Cluster of Excellence \& Mainz Institute for Theoretical Physics, Johannes Gutenberg University, Staudingerweg 9, 55128 Mainz, Germany}

\emailAdd{prisco.lo.chiatto@mpp.mpg.de}
\emailAdd{sebastian.biekoetter@kit.edu}
\emailAdd{nils.wagner@tum.de}
\emailAdd{yu001@uni-mainz.de}

\preprint{%
	\begin{tabular}{@{}r@{}}
		MITP--26--042\\
		MPP--2026--130\\
		TUM--HEP--1612/26
	\end{tabular}
}

\abstract{The late-time ringdown dynamics of near-extremal black holes (BHs) are expected to be dominated by zero-damped modes (ZDMs), whose decay rates are parametrically suppressed relative to those of ordinary quasinormal modes. In this paper, we demonstrate that exact WKB methods provide an exceptionally powerful framework for analyzing the ZDM spectrum of near-extremal Reissner–Nordström (RN) BHs. Focusing on massless, neutral scalar modes propagating on an RN background, we present the full Stokes geometry derived from the radial eigenvalue problem and establish the corresponding exact quantization condition (EQC). Our analytic computation of the Voros symbols entering the EQC achieves higher-order accuracy for the ZDM spectrum compared to previous studies and is systematically improvable. Ergo, this work serves as a proof of concept for investigations of ZDM spectra of other systems using exact WKB methods.}

\begin{document}
\maketitle
\flushbottom

\section{Introduction and motivation}

\noindent
Determining the response of black holes (BHs) to perturbations is a long-standing calculational challenge~\cite{Regge:1957td,Vishveshwara:1970cc, Press:1971wr, Teukolsky:1973ha,Chandrasekhar:1975zza, Kokkotas:1999bd, Berti:2009kk, Konoplya:2011qq}.
For small perturbations, the dynamics are governed by a linear wave equation on the BH background, and the corresponding retarded Green's function, through its analytic structure, encodes the different regimes of the BH response~\cite{Leaver:1986gd, Ching:1995tj, Berti:2009kk}. The prompt response is associated with direct propagation and early nonresonant scattering, while poles of the analytically continued Green's function give quasinormal modes (QNMs), which govern the resonant intermediate-time response.
QNMs are the intrinsic resonances of the BH geometry and characterize its ringdown regime, providing the basis for BH spectroscopy and ringdown modeling~\cite{Dreyer:2003bv,Berti:2005ys,Gossan:2011ha,Isi:2019aib} (see reference~\cite{Berti:2025hly} for a recent review on the ringdown phase).
In asymptotically flat spacetimes, the Green's function also has branch-cut singularities associated with long-range backscattering, leading to the Price tails that dominate at asymptotically late times~\cite{Price:1971fb, Price:1972pw, Ching:1995tj}.\footnote{Note that this split of the time evolution dynamics into prompt response, ringdown, and non-exponential tail is structurally analogous to the time evolution of unstable quantum states~\cite{TimeEvolutionQuantumSystem,FondaDecayTheory,PeresDecayLaw}.}\\

\noindent 
BHs are characterized by mass, electric charge, and angular momentum, which completely determine the metric. Importantly, distinct horizons become degenerate when charges take on {\it extremal} values. In fact, a smooth stationary extremal BH has a degenerate Killing horizon characterized by vanishing surface gravity and temperature, and hence exhibits a universal near-horizon scaling limit with enhanced symmetry~\cite{Gibbons:1987ps, Bardeen:1999px, Kim:2001ev, Kim:2012mh, Kunduri:2013gce}. In this limit, the near-horizon geometry develops a long throat that effectively decouples from the asymptotic spacetime. This emergent separation of scales is reflected directly in the QNM spectrum. Close to extremality, the spectrum generally separates into two sectors, each carrying information about one of the two relevant dimensionful scales, namely the BH mass and its temperature~\cite{Detweiler:1980gk, Yang:2012pj, Yang:2013uba, Konoplya:2013rxa, Richartz:2014jla, Cook:2014cta, Dias:2015wqa,  Zimmerman:2015trm,Davey:2023fin}. Conventional damped modes have frequencies and damping rates of order the inverse BH mass.
On the other hand, zero-damped modes (ZDMs) have damping rates proportional to the BH temperature and therefore become arbitrarily long-lived as extremality is approached.  \\

\noindent
We remark that ZDMs have been objects of intense interest in the literature. For instance, in the phenomenologically most relevant example, namely rapidly rotating Kerr BHs, ZDMs may affect the late stages of ringdown through their collective response. Moreover, as extremality is approached, they accumulate at a distinguished frequency: at zero frequency for neutral massless perturbations, and at the electrostatic threshold for charged fields~\cite{Yang:2012pj,Yang:2013uba,Zimmerman:2015trm}.
Interestingly, this accumulation is closely related to the nonuniform nature of the extremal and late-time limits and provides a frequency-domain perspective on the anomalous horizon behavior of extremal BHs, including the derivative instability uncovered by Aretakis~\cite{Aretakis:2011ha,Lucietti:2012xr,Casals:2016mel,Gelles:2025gxi}.
Separately, long-lived near-extremal modes have also been invoked in proposed connections between BH relaxation and the weak gravity conjecture~\cite{Hod:2017uqc,Urbano:2018kax,Harlow:2022ich}, and in discussions of relaxation bounds and BH thermodynamics~\cite{Gruzinov:2007ai,Hod:2007tb,Hod:2008zz,Hod:2010hw}.\\

\noindent 
The analytic structure of the linear wave equation plays a pivotal role in understanding the distinction between the usual damped modes and the ZDMs. 
Away from extremality, the outer and inner horizons are distinct regular singular points of the underlying equation. 
In contrast, in the extremal limit they coalesce into a degenerate horizon, and the equation undergoes a confluent limit~\cite{Zimmerman:2015trm}.\footnote{This confluent structure explains why a na\"{\i}ve local WKB expansion about the usual exterior photon-sphere saddle~\cite{Schutz:1985km,Iyer:1986np,Iyer:1986nq,Cardoso:2008bp} is not sufficient for the zero-damped family.} To resolve the ZDM spectrum accordingly, it is useful to take a double-scaling limit, in which the radial coordinate is rescaled such as to parameterize the region between the nearly degenerate horizons.
In this scaling, the near-horizon throat remains finite, and the zero-damped spectrum survives as a nontrivial resonance problem. \\

\noindent 
We remark that BH QNMs have been computed using, among other techniques, continued fractions~\cite{Leaver:1985ax,Leaver:1990zz}, local Wentzel--Kramers--Brillouin (WKB) expansions~\cite{Schutz:1985km,Iyer:1986np,Iyer:1986nq,Konoplya:2019hlu},
complex-plane monodromy methods \cite{Motl:2002hd,Motl:2003cd,Andersson:2003fh}, and Borel-resummed perturbative expansions \cite{Hatsuda:2019eoj,Eniceicu:2019npi}. More recently, exact WKB methods have been applied to BH perturbation problems~\cite{Miyachi:2025ptm,Miyachi:2025dyk,Saxena:2026zme,Hatsuda:2026ghx}.\footnote{Related EQCs for BH QNMs have also been formulated in terms of quantum Seiberg--Witten periods~\cite{Aminov:2020yma}.} However, their application to the ZDM sector remains unexplored. \\

\noindent 
The Reissner--Nordstr\"om (RN) geometry provides a particularly clean setting in which to analyze the characteristic features of ZDM spectra. 
Although astrophysical BHs are expected to carry negligible electric charge, RN BHs possess the essential two-horizon structure and near-extremal throat, where spherical symmetry reduces the perturbation problem to a radial ordinary differential equation (ODE).  For an extremal RN BH, the throat is described by $\mathrm{AdS}_2\times S^2$~\cite{Bertotti:1959pf,Robinson:1959ev,Kim:2012mh,Kunduri:2013gce}, and, near extremality, the ZDM family can, at leading order, be understood as a set of resonances in this geometry.
This allows the near-horizon geometry to be disentangled from additional complications, such as frequency-dependent angular separation constants in rotating (Kerr) geometries. \\

\noindent
In this work, we develop an exact WKB treatment of a neutral massless scalar in the near-extremal RN geometry.
We work in the regime appropriate to the zero-damped branch, derive the associated quantization condition, and compare the resulting ZDM spectrum to previous analytic calculations~\cite{Kim:2001ev,Hod:2010hw,Chen:2012zn,Kim:2012mh,Eniceicu:2019npi}.
We validate our new higher-order corrections against numerical determinations using Leaver's continued-fraction method~\cite{Leaver:1985ax, Leaver:1990zz, Berti:2009kk, Konoplya:2011qq}. \\

\noindent 
Our paper is organized as follows. In section~\ref{sec:Introduction_Exact_WKB}, we provide a brief review of the exact WKB method and introduce our notation for the connection formulae.
In section~\ref{sec:Boundar_Value_Problem_Setup}, we present the scenario of the near-extremal RN BH, and discuss the reduction of the Klein--Gordon equation in the RN BH background to a one-dimensional second-order ODE, following Teukolsky~\cite{Teukolsky:1973ha}.
We also discuss the boundary conditions that render the ODE into an eigenvalue problem.
In section~\ref{sec:StokesEQC}, we present the Stokes geometry and the quantization condition of the eigenvalue problem from section~\ref{sec:Boundar_Value_Problem_Setup}.
In section~\ref{sec:Spectrum_Computation}, we calculate the Voros symbols that enter the quantization condition, determine the QNM spectrum, and perform both analytic and numerical cross-checks.
We conclude in section~\ref{sec:Outlook}.

\section{Basics of exact WKB}
\label{sec:Introduction_Exact_WKB}

Let us briefly review the exact WKB method to establish our conventions and introduce our notation. Readers familiar with the framework may skip this section, while we aim to make the essential ideas accessible to those unfamiliar with the subject. For a more comprehensive introduction, we refer to the many excellent resources available in the literature, such as the authoritative textbook by Kawai and Takei~\cite{KT05}.

\subsection{Riccati equation and formal WKB solutions}

We begin with the second-order ODE
\begin{align}
	\bigg\{\!\!\!\:-\!\!\:\frac{\mathrm{d}^2}{\mathrm{d}z^2}+\eta^2 Q(z)\bigg\}\:\!\Psi(z,\eta)=0\, ,
	\label{eq:SecondOrderODE}
\end{align}
where $\eta>0$ is a real expansion parameter, typically assumed to be large. Introducing the exponential ansatz $\Psi(z,\eta)=\exp\:\!\cbig\{\int^z S(z',\eta)\,\mathrm{d}z'\cbig\}$ transforms the second-order ODE~\eqref{eq:SecondOrderODE} into the Riccati equation
\begin{align}
	S(z,\eta)^2+S'(z,\eta) = \eta^2 Q(z) \, .
	\label{eq:Riccati_Equation}
\end{align}
Assuming the potential $Q(z)$ to be independent of $\eta$,\footnote{Identical considerations apply in case $Q(z,\eta)$ admits an expansion in inverse powers of $\eta$~\cite{KT05}.} this equation is solved order by order in a large-$\eta$ expansion by formally setting
\begin{align}
	S(z,\eta)=\!\sum_{k=-1}^\infty \eta^{-k}S_k(z)\, , 
	\label{eq:SSeries}
\end{align}
leading to the relation 
\begin{align}
	\mathlarger{\sum}_{k\,=\,-2}^\infty \eta^{-k} \!\! \mathlarger{\sum}_{\substack{n\:\!+\:\!m\,=\,k\\[0.05cm] n,m\,\geq\,-\:\!1}} \!\! S_n(z)S_m(z) + \!\mathlarger{\sum}_{k\,=\,-1}^\infty \eta^{-k}S_k'(z) = \eta^2 Q(z) \, .
\end{align}
Equating powers of $\eta$ yields two formal solutions $S^{(\pm)}(z,\eta)$, distinguished by the choice of branch of the square root,
\begin{align}
	S_{-1}^{(\pm)}(z)=\pm \sqrt{Q(z)}\: , 
    \label{eq:LO_WKB}
\end{align}
with the remaining coefficients determined recursively by
\begin{align}
	\mathlarger{\sum}_{\substack{n\:\!+\:\!m\,=\,k\\[0.05cm] n,m\,\geq\,-\:\!1}} \!\! S_n(z)S_m(z) + S_k'(z)= 0 \, .
	\label{eq:Recursion}
\end{align}
Although the Riccati equation~\eqref{eq:Riccati_Equation} is of first order, its two formal WKB branches arising from the sign choice in the leading-order coefficient~\eqref{eq:LO_WKB} give the two linearly independent solutions of equation~\eqref{eq:SecondOrderODE}. The two branches $S^{(\pm)}$ are naturally decomposed into parts that are even or odd under the exchange $S^{(+)}\leftrightarrow S^{(-)}$, granting the definitions
\begin{subequations}
    \begin{align}
	S_\mathrm{even}(z,\eta)&=\frac{S^{(+)}(z,\eta)+S^{(-)}(z,\eta)}{2}=\sum_{k=0}^\infty \eta^{-2k}S_{2k}(z)\, ,  \\ 
    S_\mathrm{odd}(z,\eta)&=\frac{S^{(+)}(z,\eta)-S^{(-)}(z,\eta)}{2}=\!\sum_{k=-1}^\infty \eta^{-(2k+1)}S_{2k+1}(z)\, .
\end{align}
\end{subequations}
The Riccati equation then implies
\begin{align}
	S_\mathrm{even}(z,\eta) = -\frac{1}{2}\frac{\mathrm{d}}{\mathrm{d}z}\log\!\big[S_\mathrm{odd}(z,\eta)\big]\, ,
	\label{eq:Seven}
\end{align}
which allows the formal WKB solutions $\Psi_\pm^{(\mathrm{formal})}(z,\eta)$ to be written in the compact form
\begin{align}
	\Psi_\pm^{(\mathrm{formal})}(z,\eta) = \frac{1}{\sqrt{S_\mathrm{odd}(z,\eta)}}\:\!  \exp\cbigg\{\pm\int^z S_\mathrm{odd}(z',\eta)\,\mathrm{d}z'\cbigg\}\, ,
	\label{eq:FormalSol}
\end{align}
where the sign coincides with the sign chosen in equation~\eqref{eq:LO_WKB}.

\subsection{Stokes geometry and Voros symbols}

At this stage, the formal WKB solutions~\eqref{eq:FormalSol} obtained from the Riccati equation are defined only locally in $z$. Their global analytic continuation is governed by the zeros and singularities of $Q(z)$. Zeros of $Q(z)$ are of particular importance and are referred to as \emph{turning points}. In the usual quantum-mechanical setting, real turning points separate classically allowed and forbidden regions, where the leading WKB solutions are oscillatory and exponential, respectively. More generally, also in the complex domain, turning points play a distinguished role in the local WKB analysis, since the two leading Riccati branches~\eqref{eq:LO_WKB} coalesce there. Poles of $Q(z)$, by contrast, correspond to singular points of the differential equation~\eqref{eq:SecondOrderODE}.\\

\begin{figure}[h]
	\centering
	\begin{tikzpicture}

		\draw[very thick,red] (0,0) -- (2,0);
		\draw[very thick,red] (0,0) -- (120:2);
		\draw[very thick,red] (0,0) -- (-120:2);
		\draw[decorate,decoration={zigzag,amplitude=0.75mm,segment length=1.75mm},line join=round,line cap=round,rounded corners=0.45mm, very thick,black] (0,0) -- (-2,0);

		\filldraw[very thick, fill=black] (-1/8,{-1/8*sqrt(3)/3})-- (1/8,{-1/8*sqrt(3)/3})-- (0,{1/4*sqrt(3)/3})-- cycle;

		\begin{scope}[xshift=2cm,yshift=0cm]
			\filldraw[fill=gray!15,draw=gray!20!black,semithick] (0,0) circle[radius=0.15cm];
			\draw (0,0) node[scale=0.7] {$\bm{+}$};
		\end{scope}

		\begin{scope}[shift={(120:2cm)}]
			\filldraw[fill=gray!15,draw=gray!20!black,semithick] (0,0) circle[radius=0.15cm];
			\draw (0,0) node[scale=0.7] {$\bm{-}$};
		\end{scope}

		\begin{scope}[shift={(-120:2cm)}]
			\filldraw[fill=gray!15,draw=gray!20!black,semithick] (0,0) circle[radius=0.15cm];
			\draw (0,0) node[scale=0.7] {$\bm{-}$};
		\end{scope}

		\begin{scope}[xshift=7cm]
			\draw[very thick,red] (0,0) -- (2,0);
			\draw[very thick,red] (0,0) -- (120:2);
			\draw[very thick,red] (0,0) -- (-120:2);
			\draw[decorate,decoration={zigzag,amplitude=0.75mm,segment length=1.75mm},line join=round,line cap=round,rounded corners=0.45mm, very thick,black] (0,0) -- (-2,0);

			\filldraw[very thick, fill=black] (-1/8,{-1/8*sqrt(3)/3})-- (1/8,{-1/8*sqrt(3)/3})-- (0,{1/4*sqrt(3)/3})-- cycle;

			\begin{scope}[xshift=2cm,yshift=0cm]
				\filldraw[fill=gray!15,draw=gray!20!black,semithick] (0,0) circle[radius=0.15cm];
				\draw (0,0) node[scale=0.7] {$\bm{-}$};
			\end{scope}

			\begin{scope}[shift={(120:2cm)}]
				\filldraw[fill=gray!15,draw=gray!20!black,semithick] (0,0) circle[radius=0.15cm];
				\draw (0,0) node[scale=0.7] {$\bm{+}$};
			\end{scope}

			\begin{scope}[shift={(-120:2cm)}]
				\filldraw[fill=gray!15,draw=gray!20!black,semithick] (0,0) circle[radius=0.15cm];
				\draw (0,0) node[scale=0.7] {$\bm{+}$};
			\end{scope}
		\end{scope}

	\end{tikzpicture}
	\caption{Local structure around a simple turning point $z_\mathrm{turn}$, represented by a filled triangle. Three Stokes curves (red, solid) emanate from the turning point, together with a branch cut (black, zigzag) fixing the overall branch. The sign associated with each Stokes curve indicates which of the two WKB solutions $\Psi_\pm$ is dominant along that curve.}
	\label{fig:SimpleTP}
\end{figure}
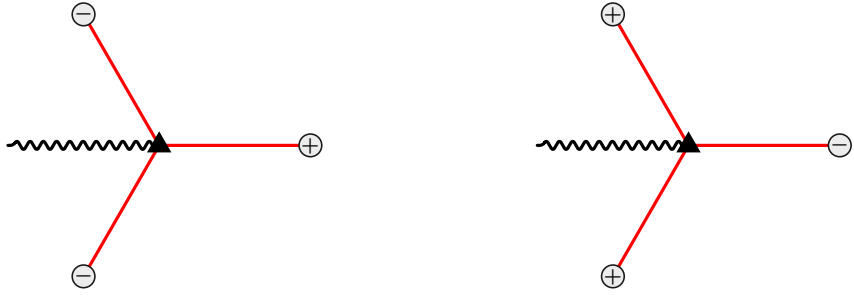

\noindent 
A \emph{Stokes curve} emanating from a turning point $z_\mathrm{turn}$ is defined by\footnote{Note that in parts of the literature, the terms \emph{Stokes curve} and \emph{anti-Stokes curve} are interchanged. The latter, in our convention, is a line of constant real part of the same expression.}
\begin{equation}
	\mathrm{Im}\cbigg\{\int_{\scalebox{0.75}{$z_\mathrm{turn}$}}^z \sqrt{Q(z')}\,dz'
	\cbigg\} \:=0\, .
	\label{eq:StokesCurves}
\end{equation}
From each \emph{simple} turning point, characterized by $Q'(z_\mathrm{turn})\neq 0$, three Stokes curves emanate. A branch cut additionally has to be introduced to fix a branch of $\sqrt{Q(z)}$, as illustrated in figure~\ref{fig:SimpleTP}. We label each curve by the sign of
\begin{equation}
	\mathrm{Re}\cbigg\{\int_{\scalebox{0.75}{$z_\mathrm{turn}$}}^z \sqrt{Q(z')}\,dz' 
	\cbigg\} \, ,
\end{equation}
indicating which of the WKB solutions $\Psi_\pm$ is dominant along it. The network of Stokes curves constitutes the \emph{Stokes graph} and divides the complex $z$-plane into \emph{Stokes regions}.\\

\noindent 
Since the original series expansion~\eqref{eq:SSeries} is generally a divergent asymptotic series, the formal WKB solutions~\eqref{eq:FormalSol} must be resummed to yield genuine solutions of the differential equation~\eqref{eq:SecondOrderODE}. For a non-degenerate Stokes graph (see section~\ref{sec:CriticalStokesGraph}) and under moderate analyticity assumptions on $Q(z)$, the WKB solutions are Borel summable within each Stokes region, where the Borel sums yield analytic solutions of equation~\eqref{eq:SecondOrderODE}~\cite{Vor83,AKT91,KT05}. We will distinguish the formal WKB solutions~\eqref{eq:FormalSol} from their Borel-resummed counterparts $\Psi_\pm(z,\eta)$ by dropping the superscript ``formal'', setting
\begin{equation}
    \Psi_\pm(z,\eta) = \mathcal{S}\Psi_\pm^{(\mathrm{formal})}(z,\eta)\, ,
\end{equation}
where $\mathcal{S}$ denotes Borel resummation.\\

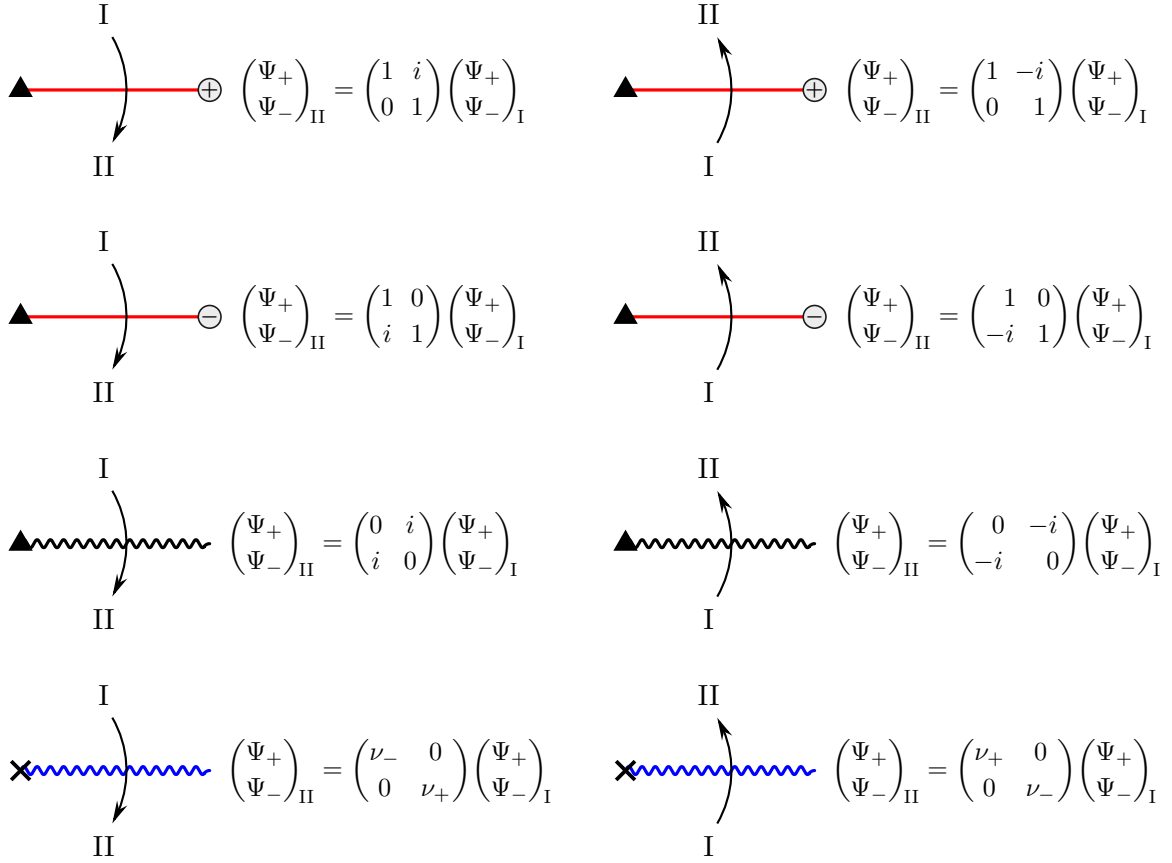
\begin{figure}[h]
	\centering
	\begin{tikzpicture}

		\begin{scope}[xshift=8cm,yshift=0cm]
			\draw[very thick,red] (0,0) -- (2.5,0);

			\filldraw[very thick, fill=black] (-1/8,{-1/8*sqrt(3)/3})-- (1/8,{-1/8*sqrt(3)/3})-- (0,{1/4*sqrt(3)/3})-- cycle;

			\draw[-{Stealth[length=0.3cm,width=0.15cm,flex]}, thick] (-30:1.4cm) arc[start angle=-30, end angle=30, radius=1.4cm];

			\begin{scope}[xshift=2.5cm,yshift=0cm]
				\filldraw[fill=gray!15,draw=gray!20!black,semithick] (0,0) circle[radius=0.15cm];
				\draw (0,0) node[scale=0.7] {$\bm{+}$};
			\end{scope}

			\draw (1.1,-1) node[scale=1] {I};
			\draw (1.1,1) node[scale=1] {II};

			\draw (2.75,0) node[anchor=west,scale=0.9] {$\displaystyle{\bigg(\begin{matrix} \Psi_+\\ \Psi_-\end{matrix}\bigg)_{\!\mathrm{II}} = \bigg(\begin{matrix} 1 & \,-i\:\! \\ 0 & \,\phantom{+}1\:\! \end{matrix}\bigg) \bigg(\begin{matrix} \Psi_+ \\ \Psi_-\end{matrix}\bigg)_{\!\mathrm{I}}}$};
		\end{scope}

		\begin{scope}[xshift=8cm,yshift=-3cm]
			\draw[very thick,red] (0,0) -- (2.5,0);

			\filldraw[very thick, fill=black] (-1/8,{-1/8*sqrt(3)/3})-- (1/8,{-1/8*sqrt(3)/3})-- (0,{1/4*sqrt(3)/3})-- cycle;

			\draw[-{Stealth[length=0.3cm,width=0.15cm,flex]}, thick] (-30:1.4cm) arc[start angle=-30, end angle=30, radius=1.4cm];

			\begin{scope}[xshift=2.5cm,yshift=0cm]
				\filldraw[fill=gray!15,draw=gray!20!black,semithick] (0,0) circle[radius=0.15cm];
				\draw (0,0) node[scale=0.7] {$\bm{-}$};
			\end{scope}

			\draw (1.1,-1) node[scale=1] {I};
			\draw (1.1,1) node[scale=1] {II};

			\draw (2.75,0) node[anchor=west,scale=0.9] {$\displaystyle{\bigg(\begin{matrix} \Psi_+\\ \Psi_-\end{matrix}\bigg)_{\!\mathrm{II}} = \bigg(\begin{matrix} \!\!\:\phantom{+}1 & \;\,0\:\! \\ \!\!\:-i & \;\,1\:\! \end{matrix}\bigg) \bigg(\begin{matrix} \Psi_+ \\ \Psi_-\end{matrix}\bigg)_{\!\mathrm{I}}}$};
		\end{scope}

		\begin{scope}[xshift=0cm,yshift=0cm]
			\draw[very thick,red] (0,0) -- (2.5,0);

			\filldraw[very thick, fill=black] (-1/8,{-1/8*sqrt(3)/3})-- (1/8,{-1/8*sqrt(3)/3})-- (0,{1/4*sqrt(3)/3})-- cycle;

			\draw[{Stealth[length=0.3cm,width=0.15cm,flex]}-, thick] (-30:1.4cm) arc[start angle=-30, end angle=30, radius=1.4cm];

			\begin{scope}[xshift=2.5cm,yshift=0cm]
				\filldraw[fill=gray!15,draw=gray!20!black,semithick] (0,0) circle[radius=0.15cm];
				\draw (0,0) node[scale=0.7] {$\bm{+}$};
			\end{scope}

			\draw (1.1,-1) node[scale=1] {II};
			\draw (1.1,1) node[scale=1] {I};

			\draw (2.75,0) node[anchor=west,scale=0.9] {$\displaystyle{\bigg(\begin{matrix} \Psi_+\\ \Psi_-\end{matrix}\bigg)_{\!\mathrm{II}} = \bigg(\begin{matrix} 1 & \;i\:\! \\ 0 & \;1\:\! \end{matrix}\bigg) \bigg(\begin{matrix} \Psi_+ \\ \Psi_-\end{matrix}\bigg)_{\!\mathrm{I}}}$};
		\end{scope}

		\begin{scope}[xshift=0cm,yshift=-3cm]
			\draw[very thick,red] (0,0) -- (2.5,0);

			\filldraw[very thick, fill=black] (-1/8,{-1/8*sqrt(3)/3})-- (1/8,{-1/8*sqrt(3)/3})-- (0,{1/4*sqrt(3)/3})-- cycle;

			\draw[{Stealth[length=0.3cm,width=0.15cm,flex]}-, thick] (-30:1.4cm) arc[start angle=-30, end angle=30, radius=1.4cm];

			\begin{scope}[xshift=2.5cm,yshift=0cm]
				\filldraw[fill=gray!15,draw=gray!20!black,semithick] (0,0) circle[radius=0.15cm];
				\draw (0,0) node[scale=0.7] {$\bm{-}$};
			\end{scope}

			\draw (1.1,-1) node[scale=1] {II};
			\draw (1.1,1) node[scale=1] {I};

			\draw (2.75,0) node[anchor=west,scale=0.9] {$\displaystyle{\bigg(\begin{matrix} \Psi_+\\ \Psi_-\end{matrix}\bigg)_{\!\mathrm{II}} = \bigg(\begin{matrix} 1 & \;0\:\! \\ i & \;1\:\! \end{matrix}\bigg) \bigg(\begin{matrix} \Psi_+ \\ \Psi_-\end{matrix}\bigg)_{\!\mathrm{I}}}$};
		\end{scope}

		\begin{scope}[yshift=-6cm]
			\begin{scope}[xshift=8cm,yshift=0cm]
				\draw[decorate,decoration={zigzag,amplitude=0.75mm,segment length=1.75mm},line join=round,line cap=round,rounded corners=0.45mm, very thick,black] (0,0) -- (2.5,0);

				\filldraw[very thick, fill=black] (-1/8,{-1/8*sqrt(3)/3})-- (1/8,{-1/8*sqrt(3)/3})-- (0,{1/4*sqrt(3)/3})-- cycle;

				\draw[-{Stealth[length=0.3cm,width=0.15cm,flex]}, thick] (-30:1.4cm) arc[start angle=-30, end angle=30, radius=1.4cm];

				\draw (1.1,-1) node[scale=1] {I};
				\draw (1.1,1) node[scale=1] {II};

				\draw (2.6,0) node[anchor=west,scale=0.9] {$\displaystyle{\bigg(\begin{matrix} \Psi_+\\ \Psi_-\end{matrix}\bigg)_{\!\mathrm{II}} = \bigg(\begin{matrix} \!\!\:\phantom{+}0 & \;\,-i\:\! \\ \!\!\:-i & \;\,\phantom{+}0\:\! \end{matrix}\bigg) \bigg(\begin{matrix} \Psi_+ \\ \Psi_-\end{matrix}\bigg)_{\!\mathrm{I}}}$};
			\end{scope}

			\begin{scope}[xshift=0cm,yshift=0cm]
				\draw[decorate,decoration={zigzag,amplitude=0.75mm,segment length=1.75mm},line join=round,line cap=round,rounded corners=0.45mm, very thick,black] (0,0) -- (2.5,0);

				\filldraw[very thick, fill=black] (-1/8,{-1/8*sqrt(3)/3})-- (1/8,{-1/8*sqrt(3)/3})-- (0,{1/4*sqrt(3)/3})-- cycle;

				\draw[{Stealth[length=0.3cm,width=0.15cm,flex]}-, thick] (-30:1.4cm) arc[start angle=-30, end angle=30, radius=1.4cm];

				\draw (1.1,-1) node[scale=1] {II};
				\draw (1.1,1) node[scale=1] {I};

				\draw (2.6,0) node[anchor=west,scale=0.9] {$\displaystyle{\bigg(\begin{matrix} \Psi_+\\ \Psi_-\end{matrix}\bigg)_{\!\mathrm{II}} = \bigg(\begin{matrix} 0 & \;\,i\:\! \\ i & \;\,0\:\! \end{matrix}\bigg) \bigg(\begin{matrix} \Psi_+ \\ \Psi_-\end{matrix}\bigg)_{\!\mathrm{I}}}$};
			\end{scope}
		\end{scope}

		\begin{scope}[yshift=-9cm]
			\begin{scope}[xshift=8cm,yshift=0cm]
				\draw[decorate,decoration={zigzag,amplitude=0.75mm,segment length=1.75mm},line join=round,line cap=round,rounded corners=0.45mm, very thick,blue] (0,0) -- (2.5,0);

				\fill[black,line width = 0.50mm] (0,0) pic[black] {cross=5pt};

				\draw[-{Stealth[length=0.3cm,width=0.15cm,flex]}, thick] (-30:1.4cm) arc[start angle=-30, end angle=30, radius=1.4cm];

				\draw (1.1,-1) node[scale=1] {I};
				\draw (1.1,1) node[scale=1] {II};

				\draw (2.6,0) node[anchor=west,scale=0.9] {$\displaystyle{\bigg(\begin{matrix} \Psi_+\\ \Psi_-\end{matrix}\bigg)_{\!\mathrm{II}} = \bigg(\begin{matrix} \nu_{+} & \;\,0\:\! \\ 0 & \;\,\nu_{-}\:\! \end{matrix}\bigg) \bigg(\begin{matrix} \Psi_+ \\ \Psi_-\end{matrix}\bigg)_{\!\mathrm{I}}}$};
			\end{scope}

			\begin{scope}[xshift=0cm,yshift=0cm]
				\draw[decorate,decoration={zigzag,amplitude=0.75mm,segment length=1.75mm},line join=round,line cap=round,rounded corners=0.45mm, very thick,blue] (0,0) -- (2.5,0);

				\fill[black,line width = 0.50mm] (0,0) pic[black] {cross=5pt};

				\draw[{Stealth[length=0.3cm,width=0.15cm,flex]}-, thick] (-30:1.4cm) arc[start angle=-30, end angle=30, radius=1.4cm];

				\draw (1.1,-1) node[scale=1] {II};
				\draw (1.1,1) node[scale=1] {I};

				\draw (2.6,0) node[anchor=west,scale=0.9] {$\displaystyle{\bigg(\begin{matrix} \Psi_+\\ \Psi_-\end{matrix}\bigg)_{\!\mathrm{II}} = \bigg(\begin{matrix} \nu_{-} & \;\,0\:\! \\ 0 & \;\,\nu_{+}\:\! \end{matrix}\bigg) \bigg(\begin{matrix} \Psi_+ \\ \Psi_-\end{matrix}\bigg)_{\!\mathrm{I}}}$};
			\end{scope}
		\end{scope}
	\end{tikzpicture}
	\caption{Comprehensive dictionary of exact WKB connection formulae for the analytic continuation of the Borel-resummed WKB basis $\Psi_\pm$ across Stokes curves and branch cuts. Since the branch cuts emanating from the turning points and double poles, represented by filled triangles and crosses respectively, serve distinct purposes in the exact WKB analysis, we highlight them using different colors. For convenience, we explicitly display the formulae for both clockwise and counter-clockwise crossings, the latter following directly from the clockwise ones by inverting the corresponding connection matrices. Note that these rules assume that the WKB bases are properly normalized at the turning point whose Stokes curve is crossed.}
	\label{fig:Connections}
\end{figure}

\noindent 
Along a Stokes curve, one of the two solutions is dominant and the other subdominant. When crossing a Stokes curve, the Borel-resummed WKB basis undergoes a \emph{Stokes phenomenon}, and the bases in the two adjacent Stokes regions are related by an exact \emph{connection formula}. For a simple turning point, these connection formulae were established through Borel analysis by Voros, Silverstone, and subsequent authors~\cite{Vor83,Sil85,AKT91,KT05}. The corresponding connection formulae across the Stokes curves are summarized in figure~\ref{fig:Connections}, which serves as a convenient reference for the exact WKB framework. Note that the formal WKB solutions~\eqref{eq:FormalSol} together with the associated resummed counterparts $\Psi_\pm(z,\eta)$ need to be provided with a lower integration bound, supplying a normalization to the solutions. For illustration, let
\begin{equation}
	\Psi_\pm^{\scalebox{0.75}{$(z_{\mathrm{turn},i})$}}(z,\eta) = \mathcal{S}\cBigg(\frac{1}{\sqrt{S_{\mathrm{odd}}(z,\eta)}} \:\!\exp\cbigg\{\pm \mathlarger{\int}_{\scalebox{0.75}{$z_{\mathrm{turn},i}$}}^z S_\mathrm{odd}(z',\eta)\,\mathrm{d}z' \cbigg\} \!\!\:\cBigg)
	\label{eq:WKBNormalizedAtA}
\end{equation}
denote the resummed WKB solutions normalized at a turning point $z_{\mathrm{turn},i}$.\footnote{The integral from a simple turning point is understood in the standard exact-WKB sense, as one half of the corresponding contour integral on the two-sheeted spectral cover.} If the same basis is instead normalized at another turning point $z_{\mathrm{turn},j}$, then the two are related by
\begin{equation}
	\Psi_\pm^{\scalebox{0.75}{$(z_{\mathrm{turn},i})$}}(z,\eta) = \mathcal{S}\exp\cbigg\{\pm \mathlarger{\int}_{\scalebox{0.75}{$z_{\mathrm{turn},i}$}}^{\scalebox{0.75}{$z_{\mathrm{turn},j}$}} S_\mathrm{odd}(z',\eta)\,\mathrm{d}z' \cbigg\}\, \Psi_\pm^{\scalebox{0.75}{$(z_{\mathrm{turn},j})$}}(z,\eta) \, .
	\label{eq:ChangeOfWKBNormalization}
\end{equation}
The connection formulae shown in figure~\ref{fig:Connections} explicitly require the resummed solutions $\Psi_\pm(z,\eta)$ to be normalized at the turning point from which the Stokes curve emanates.\\  

\noindent 
%The local connection formulae must be supplemented by the relative normalization of WKB bases associated with different turning points. 

\noindent 
The exponential factor appearing in equation~\eqref{eq:ChangeOfWKBNormalization} was first emphasized by Voros~\cite{Vor83} and thus bears his name. To define it globally, it is natural to pass to the two-sheeted spectral curve $y^2=Q(z)$ on which $\sqrt{Q(z)}$, and hence $S_{\mathrm{odd}}$, becomes single-valued. For a closed cycle $\gamma$ on this surface, the corresponding Voros symbol is defined by
\begin{align}
	\mathcal{V}_{\gamma}= \mathcal{S}\exp\cbigg\{\mathlarger{\oint}_{\gamma} \:S_\mathrm{odd}(z,\eta)\,\mathrm{d}z \cbigg\}\, .
    \label{eq:Voros_Symob_Def}
\end{align}
In particular, if the projection $\gamma_{ij}$ encircles the two turning points $z_{\mathrm{turn},i}$ and $z_{\mathrm{turn},j}$, its period encodes the relative normalization between the WKB bases normalized at $z_{\mathrm{turn},i}$ and $z_{\mathrm{turn},j}$. The Borel-resummed Voros symbols are the all-orders analogues of the familiar leading-order WKB propagation and tunneling factors, encoding the global monodromy data of the problem~\cite{Vor83,DDP93,DP99}.\\

\noindent 
Branch cuts associated with turning points require a separate treatment compared to Stokes curves. They are introduced to specify the branches entering the WKB solutions, and crossing a branch cut does not induce a Stokes jump. Such a cut interchanges the two sheets of the spectral curve and correspondingly the two WKB branches,
\begin{equation}
\Psi_+ \longleftrightarrow \Psi_- \, .
\end{equation}
In addition, the square root appearing in the WKB prefactor may contribute a constant phase, whose precise value depends on the chosen branch conventions.
Thus, apart from the relabeling of the WKB branches, only a convention-dependent phase is introduced when crossing a branch cut, and no other Stokes data. The corresponding matrices are included in figure~\ref{fig:Connections}. \\

\noindent 
Double poles of $Q(z)$ require yet a slightly different treatment, as a cut associated with a double pole $z_\mathrm{pole}$ carries additional monodromy. Suppose that $Q(z)$ has the local expansion
\begin{align}
    Q(z)=c^2 \big(z-z_\mathrm{pole}\big)^{\!\!\:-2}+\mathcal{O}\cbig[\big(z-z_\mathrm{pole}\big)^{\!\!\:-1}\:\!\cbig]\, , \qquad\quad c=\!\!\!\!\!\!\!\mathop{\mathrm{Res}}_{\scalebox{0.8}{$\;\;\;\;\;z\!=\!z_\mathrm{pole}$}} \!\!\!\!\!\!\!\:\sqrt{Q(z)}\, ,
\end{align}
then the odd part $S_\mathrm{odd}(z,\eta)$ of the Riccati solution has a simple pole, taking the form  
\begin{align}
    S_\mathrm{odd}(z,\eta) = \frac{\rho}{z-z_\mathrm{pole}}+\mathcal{O}\cbig[\big(z-z_\mathrm{pole}\big)^{\!\!\:0}\:\!\cbig]\, ,
\end{align}
with residue~\cite{KT05} 
\begin{align}
	\rho \equiv \!\!\!\!\!\!\mathop{\mathrm{Res}}_{\scalebox{0.8}{$\;\;\;\;\;z\!=\!z_\mathrm{pole}$}} \!\!\!\!\!S_\mathrm{odd}(z,\eta) = \eta \:\!c \,\sqrt{1+\frac{1}{4\eta^2c^2}} \; .
	\label{eq:Residue_Formula_EWKB}
\end{align}
Consequently, the canonically normalized WKB solutions behave as
\begin{equation}
	\Psi_\pm(z,\eta)\sim \big(z-z_\mathrm{pole}\big)^{\!\!\:\frac{1}{2}\pm \rho}\, .
\end{equation}
The common factor $\big(z-z_\mathrm{pole}\big)^{\!\!\:\frac{1}{2}}$ originates from the WKB prefactor and accounts for the $\frac{1}{2}$ appearing in the local characteristic exponents. In contrast to a turning-point cut, the cut associated with a double pole originates from the logarithmic term in the WKB action. Under a positive analytic continuation around the pole $\log\!\big(z-z_\mathrm{pole}\big)\longmapsto \log\!\big(z-z_\mathrm{pole}\big)+2\pi i$, the two solutions acquire the monodromy factors~\cite{KT05}
\begin{align}
	\nu_\pm = \exp\!\!\:\bigg\{i\pi \Big[1\pm \sqrt{4\eta^2c^2+1}\Big]\!\!\:\bigg\}\, .
	\label{eq:Characteristic_Exponents}
\end{align}
Accordingly, the corresponding connection matrix is diagonal, as shown in figure~\ref{fig:Connections}. Unlike a turning-point cut, which only exchanges the two WKB labels up to a phase, the pole cut thus carries genuine local monodromy fixed by the residue of $S_{\mathrm{odd}}(z,\eta)$.\\

\noindent 
The Stokes graph around a double pole of $Q(z)$ exhibits a characteristic logarithmic inspiral of the Stokes curves terminating at that regular singular point. Being lines of constant phase of $\scalebox{0.75}{$\displaystyle{\int^{\scalebox{1}{$z$}}}$}\sqrt{Q(z')}\,\mathrm{d}z'$, the Stokes curves wind an infinite number of times around the double pole. The importance of this logarithmic spiral for obtaining the correct QNM spectrum of BHs has recently been emphasized in reference~\cite{Miyachi:2025ptm}.\\

\noindent 
We can now proceed to construct a global solution to equation~\eqref{eq:SecondOrderODE} by analytically continuing region-specific Borel sums through the Stokes graph via the local connection matrices shown in figure~\ref{fig:Connections}.

\subsection{Boundary-value problems, quantization, and critical Stokes graphs}
\label{sec:CriticalStokesGraph}

A central application of exact WKB analyses in physics is the determination of spectra associated with boundary-value problems. Let us therefore assume that the ODE~\eqref{eq:SecondOrderODE} is supplemented by suitable boundary conditions imposed at two distinguished endpoints or asymptotic regions. As stated before, a global solution is obtained by analytically continuing the WKB basis along a path through the Stokes graph. As summarized in figure~\ref{fig:Connections}, each crossing of a Stokes curve contributes a triangular connection matrix, whereas each crossing of a branch cut contributes the corresponding sheet-exchange or monodromy matrix, while changes of normalization between different turning points contribute diagonal factors involving the corresponding Voros symbols as
\begin{align}
	\bigg(\begin{matrix} \Psi_+\\ \Psi_-\end{matrix}\bigg)^{\!\!\!\:\scalebox{0.75}{$(z_{\mathrm{turn},i})$}} = \begin{pmatrix} \mathcal{V}_{ij}^{1/2} & 0 \\ 0 & \mathcal{V}_{ij}^{-1/2}\end{pmatrix}\! \bigg(\begin{matrix} \Psi_+\\ \Psi_-\end{matrix}\bigg)^{\!\!\!\:\scalebox{0.75}{$(z_{\mathrm{turn},j})$}}\, , \qquad \mathcal{V}_{ij} \equiv \mathcal{V}_{\gamma_{ij}}\, .
\end{align}
Starting from one boundary of the problem, the boundary condition selects a linear combination of the two WKB solutions $\Psi_\pm$ in the corresponding Stokes region. Transporting this combination to the region containing the second boundary and imposing the boundary condition then yields an \emph{exact quantization condition} (EQC). Explicitly, the EQC is obtained by requiring the \emph{connection matrix} $\mathcal{M}$ to map the coefficient vector fixed by the boundary condition at boundary 1 to the coefficient vector fixed by the boundary condition at boundary 2,
\begin{align}
	\bigg(\begin{matrix} \Psi_+\\ \Psi_-\end{matrix}\bigg)^{\!\!\!\:\scalebox{0.75}{$(\text{boundary 2})$}} \overset{!}{=} \begin{pmatrix} \mathcal{M}_{11} & \mathcal{M}_{12} \\ \mathcal{M}_{21} & \mathcal{M}_{22}\end{pmatrix}\! \bigg(\begin{matrix} \Psi_+\\ \Psi_-\end{matrix}\bigg)^{\!\!\!\:\scalebox{0.75}{$\text{(boundary 1)}$}}\, .
\end{align}
Thus, the EQC is the compatibility condition ensuring that the transport encoded by $\mathcal{M}$ is consistent with the prescribed boundary data at both ends.\footnote{Equivalently, the quantization condition may be formulated in a basis-independent way by requiring the Wronskian of the two boundary-selected solutions to vanish.}\\

\begin{figure}[h]
    \centering
    \includegraphics[width=\linewidth]{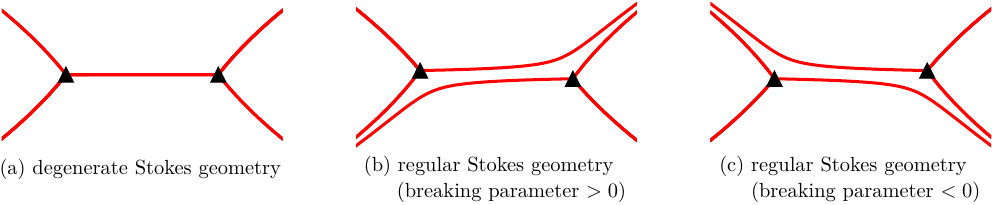}
    \caption{A degenerate Stokes geometry, featuring a regular saddle trajectory, and the two possible regularizations. While the topology of the Stokes graph, and hence the form of the EQC, depends on the regularization, the resulting spectrum does not.}
    \label{fig:DegStokesReg}
\end{figure}

\noindent 
The construction above applies directly to \emph{saddle-free} Stokes graphs, which we will also refer to informally as being \emph{non-degenerate}. If a Stokes curve emanating from a turning point either returns to the same turning point or terminates at a different one, it is called a \emph{saddle trajectory}.\footnote{A \emph{regular saddle trajectory} is a Stokes curve connecting two distinct turning points, whereas a \emph{degenerate saddle trajectory} forms a closed loop returning to the same turning point and enclosing a double pole. The changes between the two saddle-free resolutions of these configurations are conventionally referred to as a \emph{flip} and a \emph{pop}, respectively~\cite{BS15,Iwaki:2014vad}.} In the presence of such a trajectory, Borel summability generally fails in the corresponding critical direction. One therefore deforms the critical graph by introducing an infinitesimal regulator, for instance by giving $\eta$ a small phase or by deforming a parameter entering $Q(z)$, as illustrated in figure~\ref{fig:DegStokesReg}.\\

\noindent 
The Borel-resummed WKB data associated with the two possible regularizations of a critical/degenerate Stokes graph are related by a Stokes automorphism.\footnote{For regular and degenerate saddle trajectories, the corresponding wall-crossing relations are given by the Delabaere--Dillinger--Pham formula~\cite{DDP93,DP99} and the pop formula~\cite{Iwaki:2014vad}.} More generally, the corresponding wall-crossing formulae relate the exact WKB data obtained from different saddle-free resolutions of the same critical configuration. As a consequence, the EQC need not take the same algebraic form when using different regulators, but the different expressions are related by the appropriate Stokes automorphism and therefore represent the same spectral condition after analytic continuation. Thus, for a fixed boundary-value problem, the representation of the quantization condition may depend on the regulator, while the spectrum itself remains unchanged. By contrast, deformations that modify the analytic structure of the differential equation or the boundary-value problem are not necessarily related by such a wall-crossing transformation. In that case, the corresponding quantization conditions might not be related by the wall-crossing automorphisms described above, and the resulting spectra may not coincide. We will encounter examples of this distinction in the following section, beginning with the familiar harmonic oscillator.

\subsection{Example: Harmonic oscillator} 

To illustrate the exact WKB framework summarized in the previous subsection, consider the simple harmonic oscillator. We use this example to illustrate an important point: an EQC is tied to the Stokes geometry in which it is derived. If the Stokes geometry changes qualitatively as a parameter is varied, algebraic solutions of the original quantization condition do not necessarily correspond to eigenvalues of the desired boundary-value problem. To this end, let us consider the simple eigenvalue problem 
\begin{align}
	\bigg\{\!\!\!\:-\!\!\:\frac{\hbar^2}{2m}\frac{\mathrm{d}^2}{\mathrm{d}z^2}+\frac{m\omega^2z^2}{2}\bigg\}\:\!\Psi(z,\hbar)=E\Psi(z,\hbar) \, ,\qquad \Psi(z,\hbar)\xrightarrow{z\to \pm \infty} 0\, ,
\end{align}
which can be identified with the general form~\eqref{eq:SecondOrderODE} upon setting $\eta=\hbar^{-1}$ and $Q(z)=m^2\omega^2 z^2-2mE$. Imposing subdominance as $z\to\pm\infty$, our task is then to determine the admissible spectrum of the parameter $E$. It is of course well known that the spectrum is positive, which we will exploit in determining the Stokes geometry of the problem. \\

\begin{figure}[h]
	\centering
	\begin{subfigure}[c]{0.485\textwidth}
		\includegraphics[width=\textwidth]{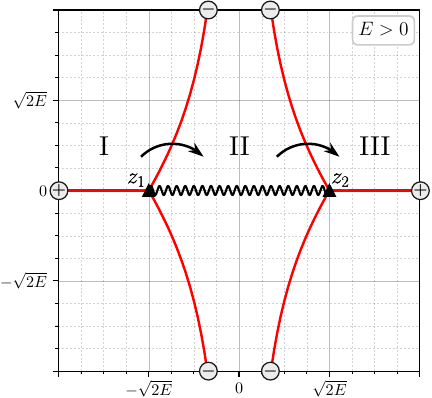}
	\end{subfigure}
	\begin{subfigure}[c]{0.485\textwidth}
		\includegraphics[width=\textwidth]{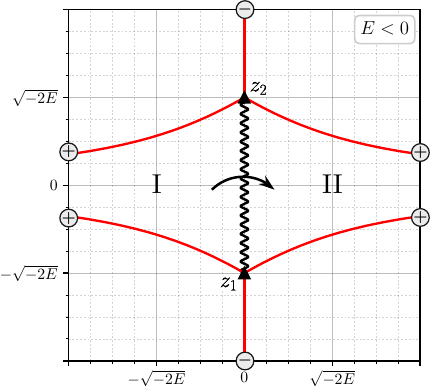}
	\end{subfigure}
	\caption{Stokes geometries arising in the exact WKB analysis of the simple harmonic oscillator for positive (\textbf{left}) and negative (\textbf{right}) values of the spectral parameter $E$. For $E>0$, the corresponding Stokes graph yields a well-defined quantization condition for the spectrum, whereas for $E<0$ no such condition arises. In both panels, we set $m\omega^2=1$ for convenience.}
	\label{fig:StokesGeometry_HarmonicOscillator}
\end{figure}

\noindent 
For $E>0$, the two turning points are 
\begin{align}
	z_{1,2} = \pm \sqrt{\frac{2E}{m\omega^2}}\: ,
\end{align}
and the associated Stokes geometry is depicted in the left panel of figure~\ref{fig:StokesGeometry_HarmonicOscillator}. As shown, we choose the branch cut to connect the two turning points $z_{1,2}$.\\

\noindent 
Normalizing the resummed WKB solutions $\Psi_\pm$ to the turning point $z_1$ and applying the rules laid out above, one finds
\begin{align}
	\bigg(\begin{matrix} \Psi_+\\ \Psi_-\end{matrix}\bigg)^{\!\!\!\:(z_1)}_{\!\mathrm{III}} &= \begin{pmatrix} \mathcal{V}_{12}^{1/2} & 0 \\ 0 & \mathcal{V}_{12}^{-1/2}\end{pmatrix}\! \bigg(\begin{matrix} \Psi_+\\ \Psi_-\end{matrix}\bigg)^{\!\!\!\:(z_2)}_{\!\mathrm{III}} \notag \\ 
	&= \begin{pmatrix} \mathcal{V}_{12}^{1/2} & 0 \\ 0 & \mathcal{V}_{12}^{-1/2}\end{pmatrix}\! \bigg(\begin{matrix} 1 & \;0\:\! \\ i & \;1\:\! \end{matrix}\bigg)\bigg(\begin{matrix} \Psi_+\\ \Psi_-\end{matrix}\bigg)^{\!\!\!\:(z_2)}_{\!\mathrm{II}} \notag\\ 
	&= \begin{pmatrix} \mathcal{V}_{12}^{1/2} & 0 \\ 0 & \mathcal{V}_{12}^{-1/2}\end{pmatrix}\! \bigg(\begin{matrix} 1 & \;0\:\! \\ i & \;1\:\! \end{matrix}\bigg)\!\begin{pmatrix} \mathcal{V}_{12}^{-1/2} & 0 \\ 0 & \mathcal{V}_{12}^{1/2}\end{pmatrix}\!\bigg(\begin{matrix} \Psi_+\\ \Psi_-\end{matrix}\bigg)^{\!\!\!\:(z_1)}_{\!\mathrm{II}} \notag \\ 
    &= \begin{pmatrix} \mathcal{V}_{12}^{1/2} & 0 \\ 0 & \mathcal{V}_{12}^{-1/2}\end{pmatrix}\! \bigg(\begin{matrix} 1 & \;0\:\! \\ i & \;1\:\! \end{matrix}\bigg)\!\begin{pmatrix} \mathcal{V}_{12}^{-1/2} & 0 \\ 0 & \mathcal{V}_{12}^{1/2}\end{pmatrix}\!\bigg(\begin{matrix} 1 & \;0\:\! \\ i & \;1\:\! \end{matrix}\bigg)\bigg(\begin{matrix} \Psi_+\\ \Psi_-\end{matrix}\bigg)^{\!\!\!\:(z_1)}_{\!\mathrm{I}} \notag \\ 
	&= \begin{pmatrix} 1 & 0 \\ i\big(1+\mathcal{V}_{12}^{-1}\big) & 1\end{pmatrix}\! \bigg(\begin{matrix} \Psi_+\\ \Psi_-\end{matrix}\bigg)^{\!(z_1)}_{\!\!\!\:\mathrm{I}} \, . 
	\label{eq:ConnectionFormula}
\end{align}
For clarity, we have displayed each step explicitly, with the four factors corresponding to the change of normalization $z_1\to z_2$, the first Stokes crossing, the inverse change of normalization, and the second Stokes crossing, respectively.\\

\noindent 
Given our choice of branch cut, we see that $\Psi_+$ constitutes the dominant solution in both limits $z\to \pm \infty$. As we see from the above connection formula~\eqref{eq:ConnectionFormula}, a pure $\Psi_-$ solution in region III will, upon analytic continuation to region I, acquire a $\Psi_+$ component unless the off-diagonal entry of the connection matrix vanishes. Therefore, for a solution satisfying the required boundary conditions to exist, we must impose the EQC
\begin{align}
	\mathcal{V}_{12} \overset{!}{=} -1\, .
\end{align}
In the present case, the Voros symbol $\mathcal{V}_{12}$ can be computed exactly. The key observation is that, aside from the turning points $z_{1,2}$, the differential equation has no finite singularities, as its only singular point lies at complex infinity. Consequently, the cycle encircling $z_{1}$ and $z_{2}$ may be continuously deformed into a contour encircling the irregular singular point at infinity. The corresponding period is then determined entirely by the residue of $S_\mathrm{odd}(z,\eta)$ at infinity, which, after taking into account the orientation of the contour, yields
\begin{align}
	\mathcal{V}_{12} = \exp\cbigg\{\mathlarger{\oint}_{\gamma_{1\!\!\;2}} S_\mathrm{odd}(z)\,\mathrm{d}z \cbigg\} &=  \exp\cbigg\{-2\pi i \!\mathop{\mathrm{Res}}_{\scalebox{0.8}{$z\!=\!\infty$}} \!\!\:\cbig(S_\mathrm{odd}(z,\hbar)\:\mathrm{d}z\cbig) \!\!\:\cbigg\} \notag \\ 
	&= \exp\cbigg\{-\frac{2\pi i}{\hbar} \!\mathop{\mathrm{Res}}_{\scalebox{0.8}{$z\!=\!\infty$}} \!\!\:\cbig(\!\!\;\sqrt{Q(z)}\:\mathrm{d}z\cbig)  \!\!\:\cbigg\}\, .
\end{align}
In the last equality, we used the fact that all higher odd WKB terms $S_{2n+1}(z)$ for $n\geq 0$ are holomorphic at $z=\infty$ and therefore have vanishing residue there, which can be inferred directly from the Riccati equation.\footnote{Note that in the special case at hand, we cannot use relation~\eqref{eq:Residue_Formula_EWKB}, as it only applies to regular singular points, whereas $z=\infty$ for the harmonic oscillator is an irregular singular point.}\\ 

\noindent 
Using 
\begin{align}
	\mathop{\mathrm{Res}}_{\scalebox{0.8}{$z\!=\!\infty$}} \!\!\:\cbig(\!\!\;\sqrt{Q(z)}\:\mathrm{d}z\cbig) &= -\mathop{\mathrm{Res}}_{\scalebox{0.8}{$w\!=\!0$}}\cbigg(\!\!\;\frac{\sqrt{Q(1/w)}}{w^2}\,\mathrm{d}w\!\!\:\cbigg) \notag\\ 
	&= -\mathop{\mathrm{Res}}_{\scalebox{0.8}{$w\!=\!0$}}\cbigg(\!\!\;\frac{1}{w^2}\sqrt{\frac{m^2\omega^2}{w^2}-2mE}\;\mathrm{d}w\cbigg)  \notag\\ 
	&= -\mathop{\mathrm{Res}}_{\scalebox{0.8}{$w\!=\!0$}} \Bigg\{\frac{m\omega}{w^3} \bigg(1-\frac{E}{m\omega^2}\,w^2+\mathcal{O}\big(w^4\big)\!\bigg)\mathrm{d}w\!\!\: \Bigg\} = \frac{E}{\omega}\, , 
\end{align}
we arrive at the EQC
\begin{align}
	\exp\cBig(-\frac{2\pi i E}{\hbar\omega} \cBig) \, \overset{!}{=} -1\, ,
\end{align}
with solutions $E_n=\hbar\omega \big(n+\frac{1}{2}\big)$. Algebraically, the quantization condition admits all $n\in \mathbb{Z}$. However, the connection formula~\eqref{eq:ConnectionFormula} from which it was obtained assumed the $E>0$ Stokes geometry. The formal branches with $n<0$, corresponding to $E<0$, therefore require a separate analysis rather than a direct extrapolation of the same quantization condition. For $E<0$, the Stokes geometry shown in the right panel of figure~\ref{fig:StokesGeometry_HarmonicOscillator} implies that there cannot be eigenfunctions that satisfy both boundary conditions, as the analytic continuation from region I to region II would simply yield
\begin{align}
	\bigg(\begin{matrix} \Psi_+\\ \Psi_-\end{matrix}\bigg)^{\!(z_1)}_{\!\mathrm{II}} &= \bigg(\begin{matrix} 0 & \;\,i\:\! \\ i & \;\,0\:\! \end{matrix}\bigg) \bigg(\begin{matrix} \Psi_+\\ \Psi_-\end{matrix}\bigg)^{\!(z_1)}_{\!\mathrm{I}} \, . 
\end{align}
We see that there cannot exist a superposition of $\Psi_+$ and $\Psi_-$ that satisfies the desired boundary conditions for any $E<0$, which thus constrains the spectrum to $n\in \mathbb{N}_0$.

\subsection{Expressing the EQC using $S(z,\eta)$}

Although EQCs are usually expressed in terms of Voros symbols involving $S_\mathrm{odd}(z,\eta)$, it sometimes proves simpler to compute the full Riccati solution $S(z,\eta)$. Let us therefore briefly state the relation between the Voros period $\smash{\mathcal{V}_{\gamma}=\mathcal{S}\exp\cbig\{\oint_{\gamma} S_\mathrm{odd}(z,\eta)\,\mathrm{d}z\cbig\}}$ and $\smash{\mathcal{S}\exp\cbig\{\oint_{\gamma} S(z,\eta)\,\mathrm{d}z\cbig\}}$. 
Since $S_\mathrm{even}(z,\eta)=-\frac{1}{2}\partial_z \log\!\big[S_\mathrm{odd}(z,\eta)\big]$, one finds~\cite{Grassi:2021wpw}
\begin{align}
	\mathcal{V}_{\gamma}= (-1)^k\mathcal{S}\exp\cbigg\{\mathlarger{\oint}_{\gamma} \,S(z,\eta)\,\mathrm{d}z \cbigg\}\, ,
    \label{eq:Relation_Voros_S}
\end{align}
where the integer $k$ is determined by the winding of $S_\mathrm{odd}(z,\eta)$ along $\gamma$, or equivalently by the zeros and poles of $S_\mathrm{odd}(z,\eta)$ enclosed by the cycle~\cite{Grassi:2021wpw}. In the case of a cycle $\gamma_{ij}$ enclosing only two simple turning points $z_{\mathrm{turn},i}$ and $z_{\mathrm{turn},j}$, as will be relevant for our analysis, we find $k=\pm 1$. Indeed, the two simple zeros of $Q(z)$ imply that $Q(z)$ winds twice around the origin along $\gamma_{ij}$, so that the cycle integral of $\mathrm{d}\log\sqrt{Q(z)}$ is $\pm 2\pi i$, with the sign determined by the orientation of the contour. After factoring out the leading term $S_{-1}(z)=\sqrt{Q(z)}$, the subleading corrections to $S_\mathrm{odd}(z,\eta)$ contribute only a logarithmic derivative of a single-valued formal power series and therefore do not modify this winding number order by order in the WKB expansion. Consequently, the contribution from $S_\mathrm{even}(z,\eta)$ reduces to an overall factor $(-1)^k=-1$, which will prove useful in sections~\ref{sec:Voros_Symbol_12} and~\ref{sec:Voros_Symbol_23}.

\section{Boundary-value problem for BH QNMs}
\label{sec:Boundar_Value_Problem_Setup}

To study BH QNMs in a simple setting, we consider scalar perturbations on a fixed BH background. Before turning to the case of interest, namely a massless, neutral scalar field $\Phi$ propagating on a near-extremal RN background, let us first establish some general considerations.

\subsection{Schr\"{o}dinger-type ODE}
Consider a spherically symmetric gravitational background (we use the mostly-plus metric convention) of the general form
\begin{align}
	\mathrm{d}s^2=-F(r)\:\! \mathrm{d}t^2 + G(r) \:\!\mathrm{d}r^2 + H(r)\:\!\mathrm{d}\Omega^2_{d-2}\, ,
	\label{eq:General_GravBackground}
\end{align}
supplemented by a $\mathrm{U}(1)_{\mathrm{em}}$ background gauge field
$A_\mu=-K(r)\delta^t_{\:\mu}$. On this background, we study a minimally coupled scalar field $\Phi$ of mass $\mu$ and charge $q$, whose equation of motion is given by
\begin{align}
	\cbig(\nabla^\mu-iqA^\mu\cbig) \!\!\;\cbig(\nabla_{\!\mu}-iqA_\mu\cbig) \Phi = \mu^2 \Phi \, .
\end{align}
For the moment, we keep the background arbitrary. Exploiting its spherical symmetry, the scalar equation can be separated by decomposing $\Phi$ into hyperspherical harmonics (HSH) on $S^{d-2}$. To this end, we employ the ansatz
\begin{align}
	\Phi\big(t,r,\Omega\big)= e^{-i\omega t} \,\mathlarger{\sum}_{j,L} \: R_{j,\omega}(r) \, \mathcal{Y}_{j,L}(\Omega)\, ,
	\label{eq:GeneralAnsatz}
\end{align} 
where $j$ denotes the total angular momentum of the HSH, while the tuple $L=\big(\ell_1,\ldots,\ell_{d-3}\big)$ parametrizes the remaining angular-momentum quantum numbers. Given the eigenvalue $-j\big(j+d-3\big)$ of the Laplacian on $S^{d-2}$ when acting on $\mathcal{Y}_{j,L}(\Omega)$, we arrive at the radial equation
\begin{align}
	R_{j,\omega}''(r)+\chi'(r) R_{j,\omega}'(r)+G(r)\,\Bigg\{\frac{1}{F(r)}\:\!\Big[\omega-qK(r)\Big]^2-\cbigg[\mu^2+\frac{j\big(j+d-3\big)}{H(r)}\cbigg]\!\!\:\Bigg\}\, R_{j,\omega}(r)=0\, ,
	\label{eq:EOM}
\end{align}
where we have used the abbreviation (see reference~\cite{Eniceicu:2019npi})
\begin{align}
	\chi(r)= \frac{1}{2}\:\!\log\!\bigg[\frac{F(r)}{G(r)}\:\! H(r)^{d-2}\bigg]\, .
\end{align}
Defining the rescaled radial function $\Psi_{j,\omega}(r)$ as 
\begin{align}
	\Psi_{j,\omega}(r) = e^{\chi(r)/2}R_{j,\omega}(r) = \bigg[\frac{F(r)}{G(r)}\:\! H(r)^{d-2}\bigg]^{\!\!\:\frac{1}{4}}R_{j,\omega}(r)
	\label{eq:ScaledRadialFunc}
\end{align}
subsequently grants the simplified Schr\"{o}dinger-type form
\begin{align}
	\!\!\scalebox{0.97}{$\displaystyle{\Psi_{j,\omega}''(r)+\Bigg\{\frac{G(r)}{F(r)}\:\!\Big[\omega-qK(r)\Big]^2-G(r)\cbigg[\mu^2+\frac{j\big(j+d-3\big)}{H(r)}\cbigg]\:\!-\:\frac{\chi'(r)^2+2\chi''(r)}{4}\Bigg\}\, \Psi_{j,\omega}(r)=0\, .}$}
	\label{eq:EOM_simplified}
\end{align}
This equation provides the general form of the radial equation considered throughout the remainder of this work. The previously introduced frequency parameter $\omega$ is determined by imposing the appropriate physical boundary conditions, thereby turning the ODE~\eqref{eq:EOM_simplified} into an eigenvalue problem. We discuss these boundary conditions in the next subsection.

\subsection{Boundary conditions in $d=4$}
\label{sec:General_Boundary_Conditions}

To determine the boundary conditions for the radial ODE~\eqref{eq:EOM_simplified} derived above, we examine the asymptotic behavior of its solutions at spatial infinity and near the future event horizon $r_h$. Let us specialize to asymptotically flat spacetimes in $d=4$, for which the metric functions and gauge potential exhibit the limiting behavior
\begin{subequations}
	\begin{align}
		F(r)&\xrightarrow{r\to\infty} 1-\frac{2M}{r}+\mathcal{O}\big(r^{-2}\big)\, , \\
		G(r)&\xrightarrow{r\to\infty} 1+\frac{2M}{r}+\mathcal{O}\big(r^{-2}\big)\, , \\
		H(r)&\xrightarrow{r\to\infty} r^2 + \mathcal{O}(r)\, , \\
		K(r)&\xrightarrow{r\to\infty} \frac{Q}{r}+\mathcal{O}\big(r^{-2}\big)\, ,
	\end{align}
\end{subequations}
where $M$ denotes the BH mass and $Q$ its charge. We further fix the gauge such that the asymptotic value of the gauge potential vanishes, \textit{i.e.} setting $K_\infty=0$. Inserting these relations, together with the resulting asymptotic behavior $\chi(r)\xrightarrow{r\to\infty}2\log(r)+\mathcal{O}\big(r^{-1}\big)$, into the radial ODE~\eqref{eq:EOM_simplified} yields
\begin{align}
	\Psi_{j,\omega}''(r)+\Bigg\{\omega^2-\mu^2+\frac{4\omega^2M-2\mu^2M-2\omega qQ}{r}+\mathcal{O}\big(r^{-2}\big)\Bigg\}\, \Psi_{j,\omega}(r)=0\, .
\end{align}
The corresponding asymptotic solutions are therefore given by
\begin{align}
	\Psi_{j,\omega}(r) \xrightarrow{r\to\infty} \exp\cbigg\{\pm i\sqrt{\omega^2-\mu^2}\:r\pm i \,\frac{2M\omega^2-\mu^2M-\omega qQ}{\sqrt{\omega^2-\mu^2}}\:\!\log(r)\cbigg\} \, .
	\label{eq:Psi_Tortoise_asymptotic}
\end{align}
Similarly, near a non-degenerate singularity at $r=r_h$, the metric functions behave as 
\begin{subequations}
	\begin{align}
		F(r)&\xrightarrow{r\to r_h} (r-r_h)f+\mathcal{O}\cbig[(r-r_h)^{2}\cbig]\, , \\
		G(r)&\xrightarrow{r\to r_h} (r-r_h)^{-1}g+\mathcal{O}(1)\, , \\
		H(r)&\xrightarrow{r\to r_h} H_h +\mathcal{O}\big(r-r_h\big)\, , \\
		K(r)&\xrightarrow{r\to r_h} K_h +\mathcal{O}\big(r-r_h\big)\, ,
	\end{align}
\end{subequations}
yielding 
\begin{align}
	\Psi_{j,\omega}''(r)+\Bigg\{\!\!\:\bigg[\frac{g}{f}\big(\omega-qK_h\big)^{\!\!\:2}+\frac{1}{4}\bigg]\!\!\;\big(r-r_h\big)^{\!\!\:-2}+\mathcal{O}\Big[\big(r-r_h\big)^{\!\!\:-1}\Big]\!\!\:\Bigg\}\, \Psi_{j,\omega}(r)=0\, .
	\label{eq:EOM_horizon_asymptotic}
\end{align}
Thus, we find the desired leading-order asymptotic behavior to be
\begin{align}
	\Psi_{j,\omega}(r) \xrightarrow{r\to r_h} \exp\cbigg\{\!\!\:\bigg[\frac{1}{2}\pm i\big(\omega-qK_h\big)\sqrt{f^{-1}g}\bigg]\log\!\big(r-r_h\big)\!\!\:\cbigg\} \, .
	\label{eq:PsiHat_Tortoise_asymptotic_2}
\end{align}
Both limiting behaviors for $r\to \infty$ and $r\to r_h$ are most naturally formulated in terms of the generalized tortoise coordinate $r^*$, implicitly given by 
\begin{align}
	\frac{\mathrm{d}r^*}{\mathrm{d}r} = \sqrt{\frac{G(r)}{F(r)}} = \left\{\begin{matrix}
	\displaystyle{1+\frac{2M}{r}}+\mathcal{O}\big(r^{-2}\big)\, ,\qquad\;\;\;& \;\;\text{for } r\to \infty \, ,\\[0.35cm]
	\sqrt{f^{-1}g} \,\big(r-r_h\big)^{\!\!\:-1} + \mathcal{O}(1)\, ,&
	\;\;\text{for } r\to r_h\, .\,\end{matrix}\right.
\end{align}
Integrating the above equality grants the relations
\begin{align}
	r^*=\left\{\begin{matrix}
	\displaystyle{r+2M\log(r)}+\mathcal{O}(1)\, ,\;\;\;\;\;\;\;\;& \;\;\text{for } r\to \infty \; (r^*\to +\infty)\, ,\\[0.3cm]
	\sqrt{f^{-1}g} \,\log\!\big(r-r_h\big) + \mathcal{O}(1)\, ,&
	\;\;\text{for } r\to r_h \; \,(r^*\to -\infty)\, ,\end{matrix}\right.
\end{align}
which allows the asymptotic solutions to be expressed as 
\begin{subequations}
	\begin{align}
		\Psi_{j,\omega}(r) &\xrightarrow{r\to\infty\: (r^*\to +\infty)} \exp\cbigg\{\pm i\sqrt{\omega^2-\mu^2}\:r^*\pm i \,\frac{\mu^2M-\omega qQ}{\sqrt{\omega^2-\mu^2}}\:\!\log(r^*)\cbigg\} \, , \\
		\Psi_{j,\omega}(r) &\xrightarrow{r\to r_h\; (r^*\to -\infty)} \exp\cbigg\{\!\!\:\bigg[\frac{1}{2}\sqrt{g^{-1}f}\pm i\big(\omega-qK_h\big)\bigg]r^*\!\!\:\cbigg\} \, .
	\end{align}
    \label{eq:asymptoticsolns}%
\end{subequations}
For a massless scalar, the first relation simplifies considerably and reduces to the familiar asymptotic behavior commonly employed in the literature (see \textit{e.g.} reference~\cite{HodWKB}). We note that the additional factor involving the surface gravity, or equivalently the near-horizon coefficients $f$ and $g$, is usually absent in the standard formulation. This difference originates from the choice of radial normalization~\eqref{eq:ScaledRadialFunc}, as one conventionally introduces a differently rescaled radial function of the form
\begin{align}
	\widehat{\Psi}_{j,\omega}(r) = \bigg[\frac{F(r)}{G(r)}\bigg]^{\!\!\:-\frac{1}{4}} \Psi_{j,\omega}(r) \xrightarrow{r\to r_h} \,\exp\!\bigg\{\!\!\!\:-\!\!\:\frac{r^*}{2}\sqrt{g^{-1}f}\!\:\bigg\} \Psi_{j,\omega}(r)\, ,
	\label{eq:ScaledRadialFunc2}
\end{align}
for which the additional near-horizon factor is compensated by the altered scaling, while the leading asymptotic behavior at $r\to\infty$ remains unchanged. This factor is therefore entirely a consequence of the chosen radial normalization~\eqref{eq:ScaledRadialFunc} and does not affect the physical ingoing or outgoing character of the wave.\\

\noindent 
Because nothing can emerge from the future event horizon, we impose the condition on~\eqref{eq:asymptoticsolns} that the perturbation propagates only into the BH there. Conversely, as the perturbation radiates away from the BH, we impose purely outgoing boundary conditions at spatial infinity. Therefore, for $\mathrm{Re}(\omega) > 0$, QNMs are obtained by solving equation~\eqref{eq:EOM_simplified} subject to the boundary conditions
\begin{subequations}
	\begin{align}
		\Psi_{j,\omega}(r) &\xrightarrow{r\to\infty\: (r^*\to +\infty)} \exp\cbigg\{i\sqrt{\omega^2-\mu^2}\:r^*+ i \,\frac{\mu^2M-\omega qQ}{\sqrt{\omega^2-\mu^2}}\:\!\log(r^*)\cbigg\} \, , \\
		\Psi_{j,\omega}(r) &\xrightarrow{r\to r_h\; (r^*\to -\infty)} \exp\cbigg\{\!\!\:\bigg[\frac{1}{2}\sqrt{g^{-1}f}- i\big(\omega-qK_h\big)\bigg]r^*\!\!\:\cbigg\} \, .
	\end{align}
	\label{eq:QNM_Boundary_Conditions}%
\end{subequations}
Note that these boundary conditions render the differential operator acting on the eigenfunctions non-Hermitian, such that its spectrum is generally complex.

\subsection{Near-extremal RN BHs}

From now on, let us focus on an RN BH, for which the metric functions take the form
\begin{subequations}
	\begin{align}
		F(r)&=G(r)^{-1}=1-\frac{2M}{r}+\frac{Q^2}{r^2}=\cBig(1-\frac{r_-}{r}\cBig)\cBig(1-\frac{r_+}{r}\cBig)\, , \\ 
		H(r)&=r^2\, , \\ 
		K(r)&=\frac{Q}{r}\, .
	\end{align}
	\label{eq:RN_MetricFunctions}%
\end{subequations}
Here, we have introduced the two characteristic radii
\begin{align}
	r_\pm=M\pm \sqrt{M^2-Q^2}\: ,
\end{align}
corresponding to the outer event horizon $r_+$ (previously denoted by $r_h$) and the inner Cauchy horizon $r_-$. Inserting the relations~\eqref{eq:RN_MetricFunctions} into the radial ODE~\eqref{eq:EOM_simplified} and recasting it into the typical WKB form~\eqref{eq:SecondOrderODE}, we obtain  
\begin{align}
	\bigg\{\!\!\!\:-\!\!\:\frac{\mathrm{d}^2}{\mathrm{d}r^2}+\eta^2Q_\mathrm{RN}(r)\bigg\}\:\!\Psi_{j,\omega}(r)=0\, ,
	\label{eq:SecondOrderODE_RN}
\end{align}
with the function $Q_\mathrm{RN}(r)$ given by 
\begin{align}
	Q_\mathrm{RN}(r)=\frac{\mu^2 r^2}{\big(r-r_+\big)\big(r-r_-\big)}&+\frac{j(j+1)}{\big(r-r_+\big)\big(r-r_-\big)} \\ 
	&-\frac{r^4}{\big(r-r_+\big)^{\!\!\:2}\big(r-r_-\big)^{\!\!\:2}}\:\!\cBig(\omega-\frac{qQ}{r}\cBig)^{\!\!\!\:2}\:\!-\:\frac{\big(r_+-r_-\big)^{\!\!\:2}}{4\big(r-r_+\big)^{\!\!\:2}\big(r-r_-\big)^{\!\!\:2}} \; . \notag
\end{align}
In equation~\eqref{eq:SecondOrderODE_RN}, we have introduced the formal counting parameter $\eta$, which is set to unity at the end of the calculation. Until then, it serves as the expansion parameter for the exact WKB analysis. We emphasize that no truncation in $\eta$ is performed: all orders are retained, either through Borel resummation or by working directly with all-order quantities.\footnote{Note that we deliberately do not decompose the displayed $Q$ into $Q_0$ and $Q_2$ with different scaling behavior under $\eta$, as was done \textit{e.g.}~in references~\cite{Miyachi:2025ptm,Hatsuda:2026ghx}. Such a decomposition was introduced solely to streamline the treatment of the boundary conditions but is not required.
%However, motivating this choice requires essentially as much explanation as working directly with the unsplit $Q$. 
Instead, we remark that for $j=0$, the turning points of the displayed WKB potential are better behaved than those of $Q_0$, making the full $Q$ easier to analyze.} \\

\noindent 
We focus on the ZDM spectrum, which arises in the near-extremal limit $Q\to M$. In the present setting, the arising modes $\omega^{(\mathrm{ZDM})}_n$ have purely imaginary frequencies, with an imaginary part that approaches zero as extremality is reached, thereby corresponding to increasingly long-lived perturbations. Note that, even though the real part of $\omega$ vanishes, the boundary conditions~\eqref{eq:QNM_Boundary_Conditions} still remain valid by analytic continuation. To parametrize the deviation of the RN BH from extremality, we introduce a small, positive, dimensionless parameter $\varepsilon\ll 1$, defined by
\begin{align}
	\varepsilon = \sqrt{1-Q^2/M^2}>0 \quad \Longrightarrow \quad r_\pm=(1\pm \varepsilon)M\, .
    \label{eq:epsilondef}
\end{align}
This allows us to systematically expand all relevant quantities in powers of $\varepsilon$. To simplify our analysis, we focus on perturbations of a massless, neutral scalar field $\Phi$, corresponding to $\mu=q=0$. In that particular case, the leading-order, near-extremal behavior of the ZDMs is known to take the form~\cite{Kim:2001ev,Hod:2010hw,Chen:2012zn,Kim:2012mh,Eniceicu:2019npi}
\begin{align}
	\omega_n^{(\mathrm{ZDM})}\sim -i\,\frac{r_+-r_-}{2r_+^2}\,\big(n+j+1\big) = -\frac{i\varepsilon}{M}\big(n+j+1\big) \Big[1 + \mathcal{O}(\varepsilon)\Big] \, , \qquad n\in\mathbb{N}_0\, .
	\label{eq:Expected_Result}
\end{align}
It is useful to introduce the rescaled spectral parameter $\Omega = iM\varepsilon^{-1}\omega$, for which the desired result takes the more convenient form  
\begin{align}
	\Omega_n^{(\mathrm{ZDM})}=\big(n+j+1\big) \Big[1 + \mathcal{O}(\varepsilon)\Big] \, .
    \label{eq:Expected_ZDM_Spectrum}
\end{align}
This rescaling is particularly convenient, as it makes the $\varepsilon$-counting manifest, allowing us to treat $\Omega$ as a real quantity of order unity. Working at finite $\varepsilon$, this assumption restricts the range of quantum numbers for which the near-extremal expansion is reliable, as $n$ and $j$ cannot be taken parametrically large. The subsequent analysis should therefore be understood as keeping $n$ and $j$ fixed with respect to the $\varepsilon$-expansion, as the corresponding error estimates are not uniform in these quantum numbers and thus do not extend to highly excited modes or large angular momentum sectors.\\ 

\noindent 
Finally, measuring all radial distances in units of the BH mass $M$ by introducing the dimensionless coordinate $z=r/M$, we arrive at the ODE
\begin{align}
	\cBigg\{-\:\!\frac{\mathrm{d}^2}{\mathrm{d}z^2}+\eta^2\Bigg[\frac{j(j+1)}{\big(z-1-\varepsilon\big)\big(z-1+\varepsilon\big)} +\frac{\varepsilon^2\big(\Omega^2 z^4-1\big)}{\big(z-1-\varepsilon\big)^{\!\!\:2}\big(z-1+\varepsilon\big)^{\!\!\:2}}\Bigg]\!\cBigg\}\:\!\Psi_{j,\Omega,\varepsilon}(z)=0\, .
	\label{eq:RN_ODE_Reduced}
\end{align}
Specializing the general QNM boundary conditions~\eqref{eq:QNM_Boundary_Conditions} to the RN background~\eqref{eq:RN_MetricFunctions}, we require the QNM eigenfunctions to asymptotically satisfy
\begin{subequations}
	\begin{align}
		\Psi_{j,\Omega,\varepsilon}(z) &\xrightarrow{z\to\infty} z^{2\eta \varepsilon \Omega} e^{\eta \varepsilon \Omega z}\, ,
		\label{eq:Psi_Tortoise_asymptotic_RN} \\[0.15cm]
		\Psi_{j,\Omega,\varepsilon}(z) &\xrightarrow{z\to z_+} \big(z-z_+\big)^{\frac{1}{2}\scalebox{0.7}{$\displaystyle{\Big[1-\Omega\sqrt{\eta^2(1+\varepsilon)^4+\Omega^{-2}(1-\eta^2)}\,\Big]}$}} \, .
		\label{eq:PsiHat_Tortoise_asymptotic_2_RN}
	\end{align}
	\label{eq:QNM_BoundaryConditions_RN}%
\end{subequations}
Although we retain the full $\eta$-dependence for the exact WKB analysis, we will set the bookkeeping parameter $\eta=1$ at the end of the calculation to make contact with the original wave equation~\eqref{eq:EOM_simplified}.

\section{Stokes geometry and ZDM quantization condition for RN BHs}
\label{sec:StokesEQC}

Having specified the boundary-value problem through the ODE~\eqref{eq:RN_ODE_Reduced} and the associated boundary conditions~\eqref{eq:QNM_BoundaryConditions_RN}, we now apply the exact WKB framework introduced in section~\ref{sec:Introduction_Exact_WKB}. \\

\noindent 
The first step is to study the Stokes geometry in the relevant parameter regime and to derive the corresponding EQC for the ZDM spectrum. We begin by examining the turning points of the WKB potential
\begin{align}
	Q(z)=\frac{j(j+1)}{\big(z-1-\varepsilon\big)\big(z-1+\varepsilon\big)} +\frac{\varepsilon^2\big(\Omega^2 z^4-1\big)}{\big(z-1-\varepsilon\big)^{\!\!\:2}\big(z-1+\varepsilon\big)^{\!\!\:2}}\, ,
	\label{eq:WKB_Potential_RN}
\end{align}
satisfying $Q(z)=0$. The WKB potential possesses four turning points, whose behavior in the small-$\varepsilon$ limit is given by
\begin{align}
	z_{1,2}&=1\pm i\varepsilon \sqrt{\frac{\Omega^2-1}{j(j+1)}-1}- \frac{2\Omega^2\varepsilon^2}{j(j+1)} + \mathcal{O}\big(\varepsilon^3\big)\, , \\[0.15cm]
	z_{3,4}&=\pm \frac{i\sqrt{j(j+1)}}{\Omega\varepsilon}- 1\pm  \frac{i\Omega \varepsilon}{\sqrt{j(j+1)}}+ \frac{2\Omega^2\varepsilon^2}{j(j+1)} + \mathcal{O}\big(\varepsilon^3\big) \, .
\end{align}
Note that the above relations are valid under the previously stated assumption $\Omega=\mathcal{O}(1)$, together with an additional restriction $j\neq 0$. The case $j=0$, for which the four turning points reduce to $z_{1,2}=\pm\Omega^{-1/2}$ and $z_{3,4}=\pm i\Omega^{-1/2}$, must be treated separately, since the corresponding Stokes geometry differs qualitatively from that of the $j\neq 0$ case. 
We focus here on $j \neq 0$ and defer the more intricate $j=0$ case to future work.\\

\noindent
While the Stokes geometry and the associated dominance assignments are invariant under a change in the sign of $\Omega$, since the WKB potential~\eqref{eq:WKB_Potential_RN} depends only on $\Omega^2$, the Stokes geometry does depend on the magnitude of $\Omega^2$. Depending on whether $\Omega^2<j(j+1)+1$ or $\Omega^2>j(j+1)+1$, the two turning points $z_{1,2}$ are either located on the real axis or acquire a small imaginary part and move into the complex plane. As we will see, similarly to the harmonic oscillator, the resulting qualitative difference in the Stokes geometries associated with the two cases imposes an additional restriction on the ZDM spectrum.

\subsection{Boundary conditions in the $\Psi_\pm$ basis}
\label{sec:BoundaryConditions_RN_WKB}

Before deriving the relevant EQCs, let us first examine which of the two WKB solutions $\Psi_\pm$ satisfies the QNM boundary condition at each endpoint. The leading WKB solutions are 
\begin{align}
    \cbig[\Psi_\pm(z)\cbig]_\text{LO WKB} &= \frac{1}{\sqrt[4]{Q(z)}}\,\exp\cbigg\{\pm\:\!\eta \int^z \sqrt{Q(z')}\,\mathrm{d}z'\cbigg\}\, . 
\end{align}
Going forward, we choose a branch assignment for $\sqrt{Q}$ such that its asymptotic behavior is given by
\begin{subequations}
    \begin{align}
        \sqrt{Q(z)}&\xrightarrow{z\to\infty} \varepsilon\lvert \Omega\rvert + \frac{2\varepsilon\lvert \Omega\rvert}{z}+ \mathcal{O}\big(z^{-2}\big)\, , \\[0.1cm]
        \sqrt{Q(z)}&\xrightarrow{z\to z_+} \frac{\sqrt{\Omega^2(1+\varepsilon)^4-1}}{2(z-z_+)}+ \mathcal{O}\Big[\big(z-z_+\big)^{\!\!\:0}\Big]\, ,
    \end{align}
    \label{eq:Branch_Assignment}
\end{subequations}
thereby resulting in the leading-order WKB expressions
\begin{subequations}
    \begin{align}
        \cbig[\Psi_\pm(z)\cbig]_\text{LO WKB} &\xrightarrow{z\to \infty} z^{\pm 2\eta \varepsilon \lvert \Omega\rvert} e^{\pm \eta \varepsilon \lvert \Omega\rvert z} \, ,\\
        \cbig[\Psi_\pm(z)\cbig]_\text{LO WKB}&\xrightarrow{z\to z_+} \big(z-z_+\big)^{\frac{1}{2}\pm \frac{1}{2}\eta\lvert \Omega\rvert\scalebox{0.7}{$\displaystyle{\sqrt{(1+\varepsilon)^4-\Omega^{-2}}}$}} \, .
    \end{align}
\end{subequations}
Comparing these asymptotic behaviors with the required boundary conditions~\eqref{eq:QNM_BoundaryConditions_RN}, we find that the QNM boundary conditions select opposite WKB branches at the two endpoints, with the specific assignment determined by the sign of $\Omega$, as summarized in table~\ref{tab:BoundaryConditions}. For $\Omega>0$, the physical solution must be asymptotic to $\Psi_+$ at spatial infinity and, after analytic continuation through the Stokes graph, becomes proportional to $\Psi_-$ at the outer horizon. Hence, the $\Psi_+$ component at the horizon must vanish. For $\Omega <0$, the assignment is reversed.
\begin{table}[H]
	\centering
	\begin{tabular}{c| c c}
		& $z\to\infty$ & $z\to z_+$ \\ \hline
		$\Omega>0$ &  $\Psi_+$ & $\Psi_-$ \\
		$\Omega<0$ &  $\Psi_-$ & $\Psi_+$
	\end{tabular}
    \caption{QNM boundary conditions expressed in terms of the WKB solutions $\Psi_\pm$ at spatial infinity and the outer horizon $z_+$, for the two signs of $\Omega$.}
    \label{tab:BoundaryConditions}
\end{table}

\noindent 
The two signs of $\Omega$ therefore impose different conditions on the global WKB connection matrix. As we will see, these conditions exclude negative values of $\Omega$ from the spectrum, thereby ruling out linearly unstable BH modes. In order to find the desired EQCs determining the spectrum, we now distinguish between the two regimes of $\Omega^2$ introduced above.

\subsection{EQC for $\Omega^2>j(j+1)+1$}
\label{sec:Case1}

\begin{figure}[h]
    \centering
    \includegraphics[width=\linewidth]{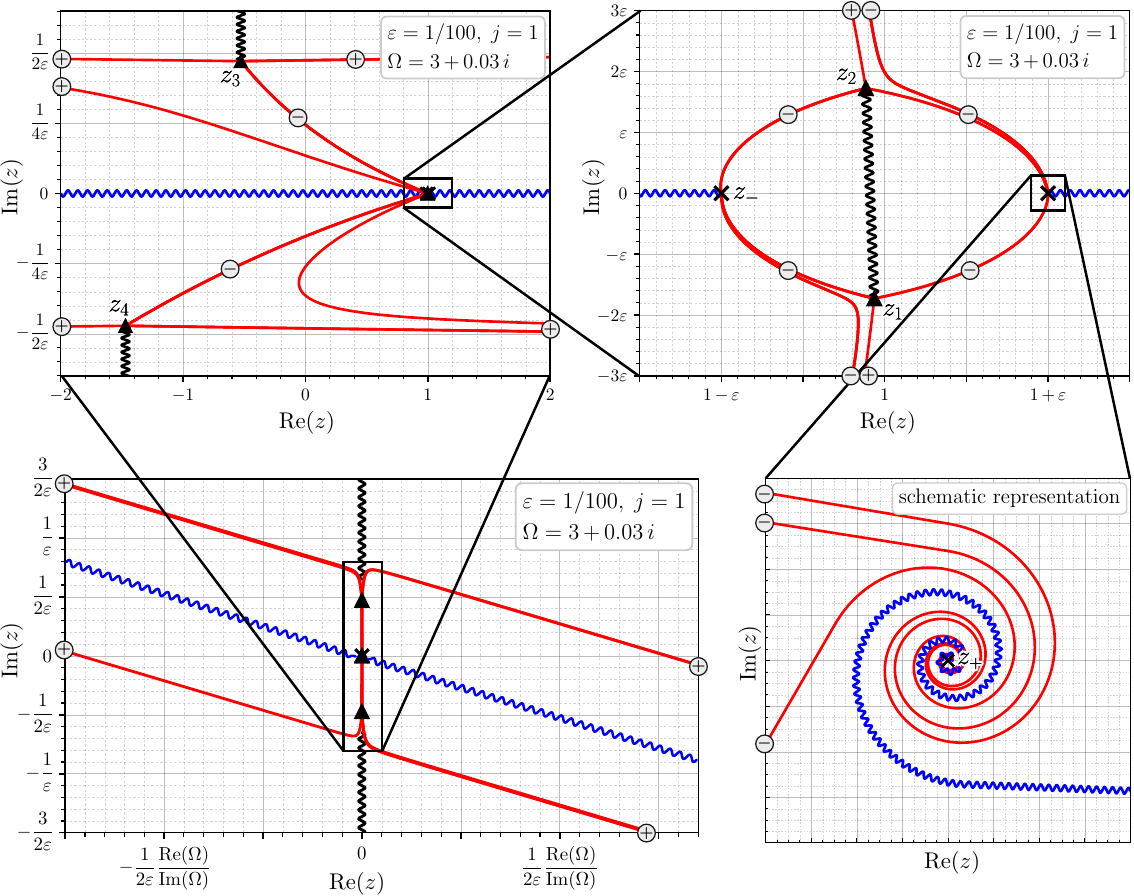}
    \caption{RN ZDM Stokes geometry for $\Omega^2>j(j+1)+1$, with the turning points $z_{1,2}$ acquiring nonzero imaginary parts. The \textbf{upper-left} panel shows the global view of the turning points, including $z_{3,4}$, while the \textbf{upper-right} panel resolves the near-horizon region $z-1=\mathcal{O}(\varepsilon)$. Since the Stokes geometry is degenerate for strictly real $\Omega$, we regulate this degeneracy by introducing a small positive imaginary part $\mathrm{Im}(\Omega)>0$. This regulator tilts the asymptotic Stokes trajectories, as displayed in the \textbf{lower-left} panel, and induces the characteristic logarithmic spiraling around the regular singular points $z_\pm$, shown schematically in the \textbf{lower-right} panel.\protect\footnotemark \;We have explicitly annotated all Stokes curves (red, solid) with their associated dominance behavior and indicated the branch cuts emanating from simple turning points (black, zigzag) and double poles (blue, zigzag).}
    \label{fig:StokesGeometry_Relevant}
\end{figure}

In the important case $\Omega^2>j(j+1)+1$, the Stokes geometry takes the form illustrated in figure~\ref{fig:StokesGeometry_Relevant}. The branch choices implied by the asymptotic behavior~\eqref{eq:Branch_Assignment}, together with the choice of a branch cut connecting the two turning points $z_1$ and $z_2$, uniquely determine the dominance assignments on all Stokes curves. In particular, the residues at the poles $z_\pm=1\pm \varepsilon$ are
\begin{align}
	c_\pm = \!\!\!\mathop{\mathrm{Res}}_{\scalebox{0.8}{$\;\;z\!=\!z_\pm$}} \!\!\sqrt{Q(z)} = \frac{1}{2}\sqrt{\Omega^2(1\pm \varepsilon)^4-1} \, ,
	\label{eq:Residues_Horizions}
\end{align}
where the fact that the two residues have the same sign follows from the chosen branch cut connecting $z_1$ and $z_2$.\footnote{The relative sign is fixed by analytic continuation across the chosen branch structure.} In the regime $\Omega^2>j(j+1)+1$, both residues are positive for sufficiently small $\varepsilon$, thereby fixing $\Psi_-$ as dominant on all Stokes curves terminating at the horizons $z_\pm$. Importantly, there is a degenerate Stokes geometry for strictly real $\Omega$, for which we introduce a small positive $\text{Im}(\Omega)>0$ to act as a regulator. \\

\footnotetext{Note that the inability to resolve the logarithmic inspiral for the parameter values used in figure~\ref{fig:StokesGeometry_Relevant} is a consequence of the strong separation of scales. Since $\mathrm{Im}(\Omega)\ll \mathrm{Re}(\Omega)$, appreciable winding occurs only in a very small neighborhood of the horizons, many orders of magnitude below the scale resolved in the upper right panel. Increasing $\mathrm{Im}(\Omega)$ relative to $\mathrm{Re}(\Omega)$ makes the spiral directly resolvable numerically and reproduces the expected logarithmic behavior. We retain a small imaginary part here because its purpose is merely to lift the degeneracy of the real-$\Omega$ Stokes graph while remaining close to that limiting geometry.}

\noindent 
%This regulator tilts the asymptotic Stokes trajectories and also introduces the characteristic logarithmic spiraling around the regular singular points $z_{\pm}$. 
As illustrated in figure~\ref{fig:StokesGeometry_Relevant}, the problem exhibits a hierarchy of scales.
Already at the level of the four turning points, two parametrically distinct regions emerge, with the near-horizon pair satisfying $z_{1,2}-1=\mathcal{O}(\varepsilon)$ and the distant pair lying at $z_{3,4}=\mathcal{O}(\varepsilon^{-1})$. 
Moreover, the regulator $\mathrm{Im}(\Omega)$ introduces a further parametric scale separation. In the limit $\mathrm{Im}(\Omega)\to 0^+$, the logarithmic spiraling around the regular singular points $z_\pm$ disappears, while the Stokes curves extending to infinity become asymptotically parallel to the real axis. For nonzero $\mathrm{Im}(\Omega)$, the Stokes curves terminating in $z_\pm$ acquire a logarithmic inspiral about each horizon, although for the regulator used here this occurs on an extremely small scale and is therefore only shown schematically. In the far-asymptotic region, the same regulator produces a small inclination of the Stokes trajectories. Although this tilt is practically invisible on the $\mathcal{O}\big(\varepsilon^{-1}\big)$ scale, its effect accumulates at large distances and generates an additional real-axis crossing of the Stokes curve emanating from $z_3$ at the parametrically larger scale $\varepsilon^{-1}\, \mathrm{Re}(\Omega)/\mathrm{Im}(\Omega)$. This additional crossing will be crucial when deriving the EQC pertaining to the present case. Resolving the complete Stokes graph therefore requires an understanding of all four parametric scales involved.\\

\begin{figure}[h]
\centering
\includegraphics[width=\textwidth]{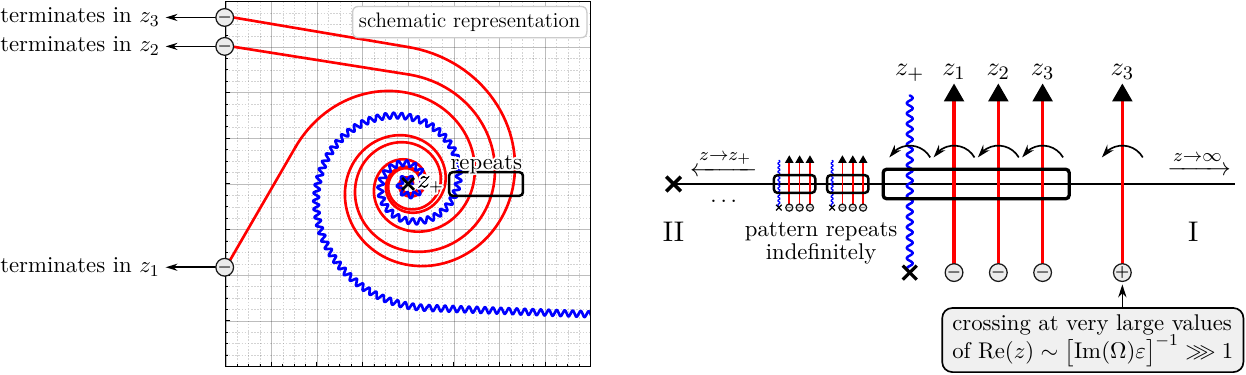}
\caption{Schematic depiction of the analytic continuation from the asymptotic region I ($z\to\infty$) inwards to the outer horizon II ($z\to z_+$). Since the Stokes curves spiral logarithmically into $z_+$, the continuation encounters an infinite sequence of crossings of both the Stokes curves and the branch cut. Note that an additional crossing occurs at extremely large values of $\mathrm{Re}(z)$ due to the infinitesimal imaginary part $\mathrm{Im}(\Omega)$ introduced to lift the degeneracy of the Stokes geometry, as described in the main text.}
\label{fig:EQC_Derivation_Illustration}
\end{figure}

\noindent 
To construct the EQC, we analytically continue the solution basis $\Psi_\pm$ from the asymptotic region $z \to\infty$ inwards to the outer horizon $z \to z_+ = 1 + \varepsilon$, as  illustrated in figure~\ref{fig:EQC_Derivation_Illustration}. As a result of the logarithmic spiraling of the Stokes curves around $z_+$, this continuation sequentially crosses the three Stokes curves emanating from $z_{1,2,3}$ and the pole branch cut an infinite number of times. \\

\noindent 
When analytically continuing the solutions from $z = \infty$ towards $z_+$, all Stokes crossings are traversed in clockwise direction, while the branch cut emanating from $z_+$ is traversed counter-clockwise. From the schematic representation in figure~\ref{fig:EQC_Derivation_Illustration}, we can directly apply the exact WKB crossing rules (see figure~\ref{fig:Connections}) to obtain the desired crossing relation. Choosing to normalize the resummed WKB solutions $\Psi_\pm$ to the turning point $z_3$, we obtain
\begin{align}
    \!\!\bigg(\begin{matrix} \Psi_+\\ \Psi_-\end{matrix}\bigg)_{\!\!\:\mathrm{II}}^{\!(z_3)} &= \lim_{N\to\infty} \cBigg\{\!\begin{pmatrix}
        \mathcal{V}_{31}^{1/2} & 0\\ 
        0 & \mathcal{V}_{31}^{-1/2}
    \end{pmatrix} \!\begin{pmatrix}
        \nu^{\scalebox{0.6}{$(z_+)$}}_+ & 0\\ 
        0 & \nu^{\scalebox{0.6}{$(z_+)$}}_-
    \end{pmatrix} \!\bigg(\begin{matrix} 1 & \;0\:\! \\ i & \;1\:\! \end{matrix}\bigg) \nonumber \\ 
    &\qquad\quad\; \times \begin{pmatrix}
        \mathcal{V}_{12}^{1/2} & 0\\ 
        0 & \mathcal{V}_{12}^{-1/2}
    \end{pmatrix} \!\bigg(\begin{matrix} 1 & \;0\:\! \\ i & \;1\:\! \end{matrix}\bigg) \!\begin{pmatrix}
        \mathcal{V}_{23}^{1/2} & 0\\ 
        0 & \mathcal{V}_{23}^{-1/2}
    \end{pmatrix} \!\bigg(\begin{matrix} 1 & \;0\:\! \\ i & \;1\:\! \end{matrix}\bigg) \!\cBigg\}^{\!\! N}\!\!\: \bigg(\begin{matrix} \:\!1 & \;i\:\! \\ \:\!0 & \;1\:\! \end{matrix}\bigg)
    \bigg(\begin{matrix} \Psi_+\\ \Psi_-\end{matrix}\bigg)_{\!\!\:\mathrm{I}}^{\!(z_3)} \\[0.2cm] 
    &= \lim_{N\to\infty} \begin{pmatrix}
        \displaystyle{\nu^{\scalebox{0.6}{$(z_+)$}}_+\sqrt{\mathcal{V}_{12}\mathcal{V}_{23}\mathcal{V}_{31}}} & \;\;0 \\[0.2cm] 
        \displaystyle{\frac{i\nu^{\scalebox{0.6}{$(z_+)$}}_-}{\sqrt{\mathcal{V}_{12}\mathcal{V}_{23}\mathcal{V}_{31}}}}\,\Big[1+\big(1+\mathcal{V}_{12}\big)\mathcal{V}_{23}\Big] & \;\;\displaystyle{\frac{\nu^{\scalebox{0.6}{$(z_+)$}}_-}{\sqrt{\mathcal{V}_{12}\mathcal{V}_{23}\mathcal{V}_{31}}}}
    \end{pmatrix}^{\!\!\! N} \!\!\:\bigg(\begin{matrix} \:\!1 & \;i\:\! \\ \:\!0 & \;1\:\! \end{matrix}\bigg)
    \bigg(\begin{matrix} \Psi_+\\ \Psi_-\end{matrix}\bigg)_{\!\!\:\mathrm{I}}^{\!(z_3)} \, . \notag 
\end{align}
Using the identity $\mathcal{V}_{12} \mathcal{V}_{23} \mathcal{V}_{31} = 1$ and defining the ratio $\mathcal{R} \equiv \nu^{\scalebox{0.6}{$(z_+)$}}_-/\nu^{\scalebox{0.6}{$(z_+)$}}_+$, the result can be brought into the convenient form
\begin{align}
    \!\bigg(\begin{matrix} \Psi_+\\ \Psi_-\end{matrix}\bigg)_{\!\!\:\mathrm{II}}^{\!(z_3)} &= \lim_{N\to\infty} \cbig(\nu^{\scalebox{0.6}{$(z_+)$}}_+\cbig)^{\!\!\:N} \begin{pmatrix}
        1 & \;\;0 \\[0.2cm] 
        \displaystyle{i\mathcal{R}\Big[1+\big(1+\mathcal{V}_{12}\big)\mathcal{V}_{23}\Big]} & \;\;\mathcal{R}
    \end{pmatrix}^{\!\!\! N} \!\!\:\bigg(\begin{matrix} \:\!1 & \;i\:\! \\ \:\!0 & \;1\:\! \end{matrix}\bigg)
    \bigg(\begin{matrix} \Psi_+\\ \Psi_-\end{matrix}\bigg)_{\!\!\:\mathrm{I}}^{\!(z_3)} \\ 
    &= \lim_{N\to\infty} \cbig(\nu^{\scalebox{0.6}{$(z_+)$}}_+\cbig)^{\!\!\:N} \begin{pmatrix}
        1 & \;\;0 \\[0.2cm] 
        \displaystyle{\frac{i\mathcal{R}\big(\mathcal{R}^N-1\big)}{\mathcal{R}-1}\Big[1+\big(1+\mathcal{V}_{12}\big)\mathcal{V}_{23}\Big]} & \;\;\mathcal{R}^N
    \end{pmatrix} \!\!\:\bigg(\begin{matrix} \:\!1 & \;i\:\! \\ \:\!0 & \;1\:\! \end{matrix}\bigg)
    \bigg(\begin{matrix} \Psi_+\\ \Psi_-\end{matrix}\bigg)_{\!\!\:\mathrm{I}}^{\!(z_3)} \, . \notag
\end{align}
Up to an overall normalization factor $\cbig(\nu^{\scalebox{0.6}{$(z_+)$}}_+\cbig)^{\!\!\:N}$, which can be absorbed into the integration constant entering the WKB functions $\Psi_\pm$, the connection matrix has the form 
\begin{align}
    \mathcal{M}&= \!\!\:\lim_{N\to\infty} \!\!\:\begin{pmatrix}
        1 & \;\;\;i\\[0.2cm] 
        \displaystyle{\frac{i\mathcal{R}\big(\mathcal{R}^N-1\big)}{\mathcal{R}-1}\Big[1+\big(1+\mathcal{V}_{12}\big)\mathcal{V}_{23}\Big]} & \;\;\;\mathcal{R}^N-\displaystyle{\frac{\mathcal{R}\big(\mathcal{R}^N-1\big)}{\mathcal{R}-1}\Big[1+\big(1+\mathcal{V}_{12}\big)\mathcal{V}_{23}\Big]}
    \end{pmatrix} .
\end{align}
Note that our discussion closely parallels the treatment presented in reference~\cite{Miyachi:2025ptm}. While traversing the logarithmic spiral yields only a triangular matrix, additional crossings outside the spiral are required to obtain a nontrivial EQC. \\

\noindent 
At this point, we recall the QNM boundary conditions from table~\ref{tab:BoundaryConditions}, which distinguish between the two possible signs of the rescaled spectral parameter $\Omega = i M \varepsilon^{-1} \omega$ and 
lead to qualitatively different conclusions. \\ 

\noindent 
First, for $\Omega < 0$, the boundary conditions require a pure $\Psi_-$ solution in region I ($z \to \infty$) to analytically continue into a pure $\Psi_+$ solution in region II ($z \to z_+$), which is not possible since 
the entries $\mathcal{M}_{11}$ and $\mathcal{M}_{12}$ are of the same order. With the connection matrix $\mathcal{M}$ and the boundary conditions specified in section~\ref{sec:BoundaryConditions_RN_WKB} being in conflict, there are no modes $\Omega<0$ that satisfy the boundary conditions.\\

\noindent 
For $\Omega>0$, the situation is more subtle. For finite $N$, it may be tempting to impose the boundary condition by simply setting the relevant matrix element $\mathcal{M}_{22}$ to zero, but this would introduce an artificial $N$-dependence and therefore does not correctly capture the  $N \to \infty$ limit. The boundary condition should instead be formulated in terms of the relative weights of the two WKB branches. Requiring a pure $\Psi_+$ solution in region I to analytically continue into a pure $\Psi_-$ solution in region II amounts to demanding
\begin{align}
    \lim_{N\to\infty} \frac{\mathcal{M}_{22}}{\mathcal{M}_{21}} = 0\, .
    \label{eq:MEratio}
\end{align}
This contrasts with the usual constraint that an individual connection matrix element vanishes, as any common, potentially divergent normalization of the WKB solutions $\Psi_\pm$ can be absorbed into their normalization factors. Explicitly, equation~\eqref{eq:MEratio} requires 
\begin{align}
    \lim_{N\to\infty} \Bigg\{1-\frac{\mathcal{R}-1}{\mathcal{R}}\frac{\mathcal{R}^N}{\mathcal{R}^N-1}\Big[1+\big(1+\mathcal{V}_{12}\big)\mathcal{V}_{23}\Big]^{-1}\Bigg\} = 0\, .
\end{align}

\noindent
To evaluate the  $N\to\infty$ limit, we first determine the magnitude of $\mathcal{R}$. Using formula~\eqref{eq:Characteristic_Exponents} for the characteristic exponents $\nu_\pm$, together with the horizon residues $c_\pm$ from equation~\eqref{eq:Residues_Horizions}, we obtain
\begin{align}
	\mathcal{R}= \exp\cBig\{-2i\pi \sqrt{\eta^2\cbig[\Omega^2(1+\varepsilon)^4-1\cbig]+1}\:\!\cBig\} \:\xrightarrow{\eta\to 1} \:\exp\cBig\{-2i\pi \Omega \big(1+\varepsilon\big)^{\!\!\:2}\cBig\} \, ,
    \label{eq:Ratio_R}
\end{align}
where we used $\mathrm{Re}(\Omega)>0$ to infer $\sqrt{\Omega^2}=\Omega$ in the last step. The small positive imaginary part $\mathrm{Im}(\Omega)>0$, introduced to lift the degeneracy of the Stokes geometry, then dictates $\lvert\mathcal{R}\rvert>1$. Consequently, the limit $N\to\infty$ amounts to replacing the factor $\mathcal{R}^N/(\mathcal{R}^N-1)$ by 1, such that the desired EQC takes the simple form
\begin{align}
    1+\mathcal{R}\big(1+\mathcal{V}_{12}\big)\mathcal{V}_{23} = 0\, .
    \label{eq:EQC_RN_Total}
\end{align}
In section~\ref{sec:Spectrum_Computation},
we will calculate the Voros symbols $\mathcal{V}_{12}$ and $\mathcal{V}_{23}$ 
and show that the above EQC precisely captures the sought-after ZDM spectrum. However, before doing so, let us briefly consider the EQC for the second case $\Omega^2 < j(j+1) + 1$.

\subsection{EQC for $\Omega^2 < j(j+1) + 1$}
\label{sec:Case2}
\begin{figure}[h]
	\centering
	\includegraphics[width=\textwidth]{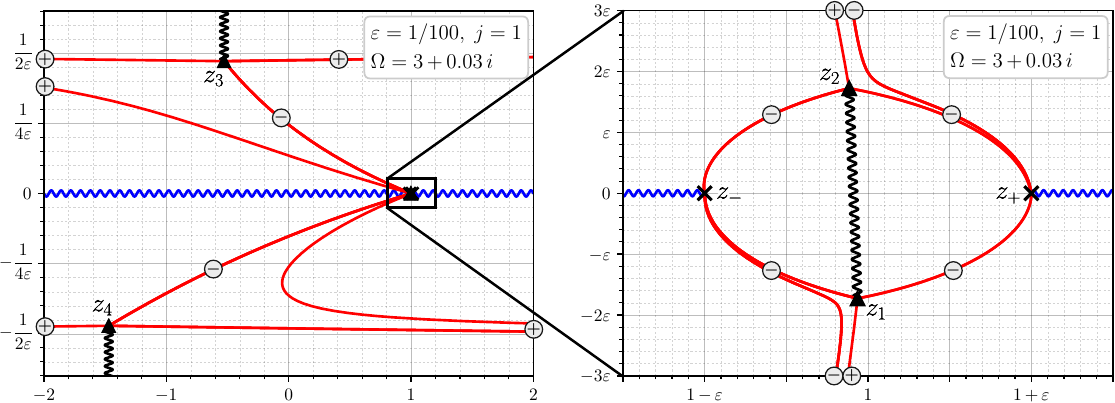}\\[0.15cm]
	\includegraphics[width=\textwidth]{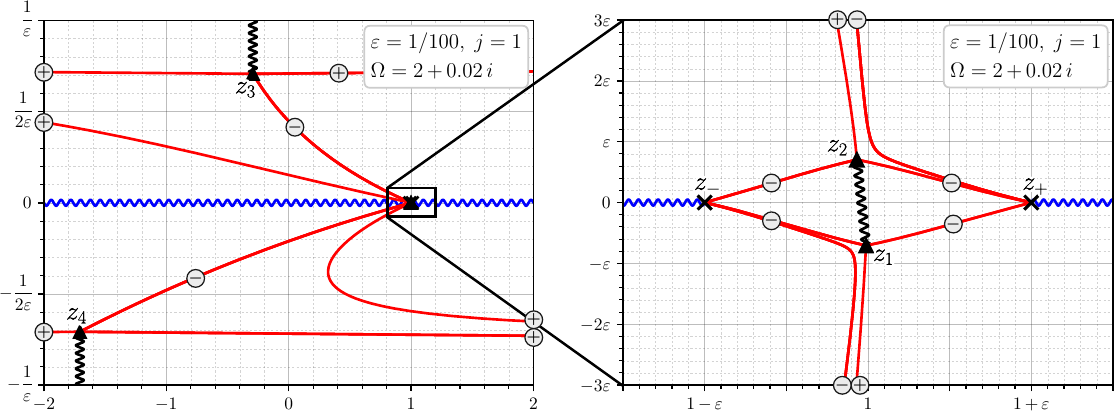}\\[0.3cm]
	\includegraphics[width=\textwidth]{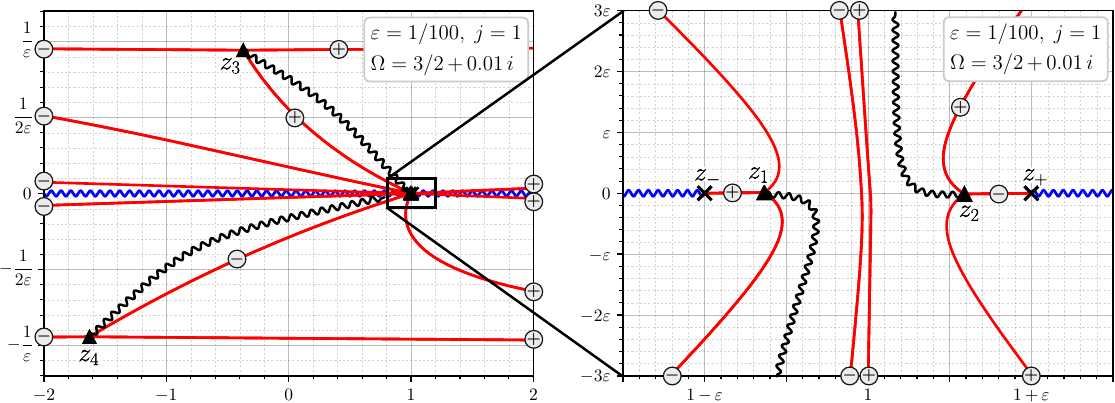}
	\caption{RN ZDM Stokes geometries for decreasing values of $\Omega$, illustrating the transition from $\Omega^2>j(j+1)+1$ to $\Omega^2<j(j+1)+1$, across which the Stokes geometry changes qualitatively. Mimicking the upper half of figure~\ref{fig:StokesGeometry_Relevant}, the \textbf{left} panels illustrate the wide view including the turning points $z_{3,4}$, whereas the \textbf{right} panels show the near-horizon region. The different \textbf{rows} correspond to different magnitudes of $\Omega^2$, decreasing from top to bottom.}
	\label{fig:StokesGeom_Change}
\end{figure}

When switching from $\Omega^2 > j(j+1)+1$ to $\Omega^2 < j(j+1)+1$, the Stokes geometry changes qualitatively, as depicted in figure~\ref{fig:StokesGeom_Change}. \\ 

\noindent 
As before, examining the region near the outer horizon $z_+$ reveals that both the branch cut and the single Stokes curve terminating at $z_+$ spiral logarithmically into the horizon in the counter-clockwise direction. The striking difference is that only a single Stokes curve terminates at $z_+$, which considerably simplifies the connection formula obtained by analytic continuation. Enforcing the branch choices dictated by the asymptotic behavior~\eqref{eq:Branch_Assignment} requires introducing branch cuts connecting $z_2$ to $z_3$ and $z_1$ to $z_4$, as shown in figure~\ref{fig:StokesGeom_Change}.\\

\begin{figure}[h]
	\centering
	\includegraphics[width=\textwidth]{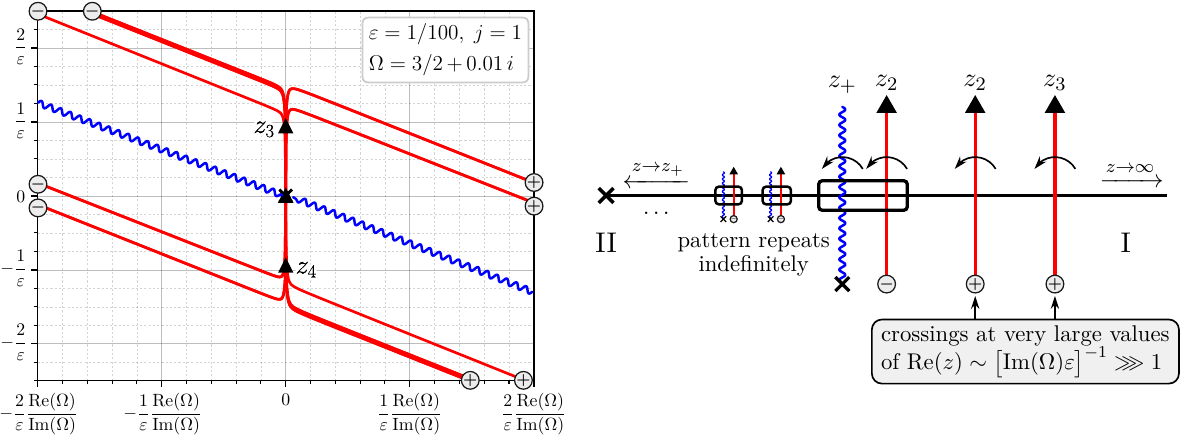}
	\caption{Far-asymptotic region of the Stokes geometry in the case $\Omega^2<j(j+1)+1$ and positive regulator $\mathrm{Im}(\Omega)>0$ (\textbf{left}), together with the illustrative depiction of the analytic continuation from the asymptotic region I ($z\to\infty$) towards the outer horizon II ($z\to z_+$) (\textbf{right}).}
	\label{fig:EQC_SmallOmega}
\end{figure}

\noindent 
In addition to the Stokes crossings arising from the spiral, there will again be additional crossings associated with the slightly tilted Stokes contours extending to infinity. The fully zoomed out view is provided in the left panel of figure~\ref{fig:EQC_SmallOmega}. Normalizing the WKB basis to $z_2$ and employing the exact WKB rules, we arrive at the connection relation
\begin{align}
	\!\bigg(\begin{matrix} \Psi_+\\ \Psi_-\end{matrix}\bigg)_{\!\!\:\mathrm{II}}^{\!\!\!\:(z_2)} &= \lim_{N\to\infty} \cBigg\{\!\begin{pmatrix}
		\nu^{\scalebox{0.6}{$(z_+)$}}_+ & 0\\ 
		0 & \nu^{\scalebox{0.6}{$(z_+)$}}_-
	\end{pmatrix} \!\!\!\:\bigg(\begin{matrix} 1 & \;0\:\! \\ i & \;1\:\! \end{matrix}\bigg) \!\cBigg\}^{\!\! N}\!\bigg(\begin{matrix} \:\!1 & \;i\:\! \\ \:\!0 & \;1\:\! \end{matrix}\bigg)\!\!\!\:\begin{pmatrix}
        \mathcal{V}_{23}^{1/2} & 0\\ 
        0 & \mathcal{V}_{23}^{-1/2}
    \end{pmatrix} \!\!\!\:\bigg(\begin{matrix} \:\!1 & \;i\:\! \\ \:\!0 & \;1\:\! \end{matrix}\bigg)\!\!\!\: \begin{pmatrix}
        \mathcal{V}_{32}^{1/2} & 0\\ 
        0 & \mathcal{V}_{32}^{-1/2}
    \end{pmatrix} \!\!\!\:
	\bigg(\begin{matrix} \Psi_+\\ \Psi_-\end{matrix}\bigg)_{\!\!\:\mathrm{I}}^{\!\!\!\:(z_2)} \nonumber \\[0.2cm] 
	&=\lim_{N\to\infty} \cbig(\nu^{\scalebox{0.6}{$(z_+)$}}_+\cbig)^{\!\!\:N} \begin{pmatrix}
		1 & \;\;0 \\[0.2cm] \displaystyle{\frac{i\mathcal{R}\big(\mathcal{R}^N-1\big)}{\mathcal{R}-1}} & \;\;\mathcal{R}^N
	\end{pmatrix} \!\begin{pmatrix}
		\:\!1 & \;\;\displaystyle{i\big(1+\mathcal{V}_{23}\big)}\:\! \\[0.2cm] 
		\:\!0 & \;\;1\:\!
	\end{pmatrix} \!
	\bigg(\begin{matrix} \Psi_+\\ \Psi_-\end{matrix}\bigg)_{\!\!\:\mathrm{I}}^{\!\!\!\:(z_2)} \, ,
\end{align}
where we used $\mathcal{V}_{32}=\mathcal{V}_{23}^{-1}$ and again set $\mathcal{R}=\nu^{\scalebox{0.6}{$(z_+)$}}_-/\nu^{\scalebox{0.6}{$(z_+)$}}_+$. Ignoring the overall prefactor $\smash{\cbig(\nu^{\scalebox{0.6}{$(z_+)$}}_+\cbig)^{\!\!\:N}}$, the connection matrix takes the form 
\begin{align}
    \mathcal{M}&= \!\!\:\lim_{N\to\infty} \!\!\:\begin{pmatrix}
    1 & \;\;\;\displaystyle{i\big(1+\mathcal{V}_{23}\big)}\\[0.2cm] 
    \displaystyle{\frac{i\mathcal{R}\big(\mathcal{R}^N-1\big)}{\mathcal{R}-1}} & \;\;\;\mathcal{R}^N-\displaystyle{\frac{\mathcal{R}\big(\mathcal{R}^N-1\big)}{\mathcal{R}-1}\big(1+\mathcal{V}_{23}\big)}
    \end{pmatrix} .
\end{align}

\noindent
As before, we perform a case-by-case analysis for positive and negative values of $\Omega$. For negative $\Omega$, we require either $\mathcal{M}_{11}=0$ or $\mathcal{M}_{12}$ to diverge in order for the ratio $\mathcal{M}_{11}/\mathcal{M}_{12}$ to vanish. The latter would require the Voros symbol $\mathcal{V}_{23}$ to diverge, which is not possible since $\mathcal{V}_{23}$ is the exponential of a finite Voros period. This again excludes $\Omega<0$ from the ZDM spectrum. \\ 

\noindent 
For positive $\Omega$, we require 
\begin{align}
    \lim_{N\to\infty} \frac{\mathcal{M}_{22}}{\mathcal{M}_{21}} = i\big(1+\mathcal{V}_{23}\big)+\frac{\mathcal{R}-1}{i\mathcal{R}}\lim_{N\to\infty} \frac{\mathcal{R}^N}{\mathcal{R}^N-1}= 0\, .
\end{align}
The characteristic exponents $\nu^{\scalebox{0.6}{$(z_+)$}}_\pm$ of $z_+$ are unchanged by the altered Stokes geometry, as the asymptotic behavior~\eqref{eq:Branch_Assignment} fixes the branch near $z_+$. Therefore, we still have $\lvert\mathcal{R}\rvert>1$, such that the limit is again 1. The corresponding EQC therefore takes the simple form
\begin{align}
   1+\mathcal{R}\mathcal{V}_{23}= 0\, .
   \label{eq:EQC_RN_Total_2}
\end{align}
We will ultimately find that, within the investigated parameter regime, this case admits no solutions for $\Omega$ and therefore does not yield any additional modes that can be identified as ZDMs.

\section{Voros symbols, ZDM spectrum, and numerical cross-checks}
\label{sec:Spectrum_Computation}

With the EQCs~\eqref{eq:EQC_RN_Total} and~\eqref{eq:EQC_RN_Total_2} in hand, we now turn to the computation of the two relevant Voros symbols $\mathcal{V}_{12}$ and $\mathcal{V}_{23}$. While $\mathcal{V}_{12}$ arises only when $\Omega^2>j(j+1)+1$, $\mathcal{V}_{23}$ enters in both EQCs. Although  $\mathcal{V}_{23}$ is formally different in the two EQCs as a result of slight differences in the branch structure and turning-point configurations, it will suffice to use the leading-order scaling of $\mathcal{V}_{23}$ with the near-extremality parameter $\varepsilon$ from equation~\eqref{eq:epsilondef}, which is identical for both cases.  This leading-order scaling is identical as it solely depends on the hierarchy of scales between the two turning points $z_2-1=\mathcal{O}(\varepsilon)$ and $z_3=\mathcal{O}\big(\varepsilon^{-1}\big)$.\\

\noindent 
After approximating the Voros symbols, the ZDM spectrum follows immediately, as the relevant EQC~\eqref{eq:EQC_RN_Total} can be systematically solved order by order in the small near-extremality parameter $\varepsilon$. In contrast, the EQC~\eqref{eq:EQC_RN_Total_2} will be found to possess no solutions in the desired parameter regime. Finally, we will compare the resulting analytical spectrum with numerical results obtained using Leaver's continued-fraction method~\cite{Leaver:1985ax, Leaver:1990zz, Berti:2009kk, Konoplya:2011qq}.

\subsection{Analysis of $\mathcal{V}_{12}$}
\label{sec:Voros_Symbol_12}

\begin{figure}[h]
	\centering
	\includegraphics[width=0.425\textwidth]{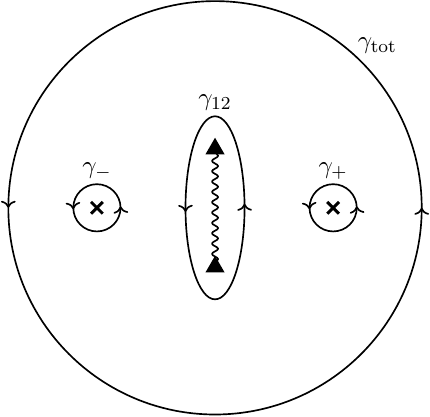}
	\caption{Schematic illustration of the integration contours at play. The Voros integral over $\gamma_{12}$ (see equation~\eqref{eq:Voros_Symob_Def}) is evaluated by considering a new circular contour $\gamma_{\mathrm{tot}}$ of $\mathcal{O}(1)$ radius and subtracting the enclosed pole contributions from $z_{\pm}$, denoted $\gamma_+$ and $\gamma_-$, separately. Note that the contour $\gamma_\mathrm{tot}$ does not encircle the distant turning points $z_{3,4}$.}
	\label{fig:Integration_Cycle_V12}
\end{figure}

For the Voros symbol $\mathcal{V}_{12}$, we work explicitly in the regime $\Omega^2 > j(j+1) + 1$, for which the Stokes geometry takes the form illustrated in figure~\ref{fig:StokesGeometry_Relevant}. We eschew a direct evaluation of the Voros symbol $\mathcal{V}_{12}$ by integrating the cycle $\gamma_{12}$, as this would be rather cumbersome. In particular, the proximity of the poles $z_\pm$, combined with the singular behavior of the higher-order terms $S_n(z)$ near the turning points $z_{1,2}$, considerably complicates a direct evaluation of equation~\eqref{eq:Voros_Symob_Def}. Instead, it is more convenient to introduce a counterclockwise circular contour $\gamma_{\mathrm{tot}}$ with radius of $\mathcal{O}(1)$ that encloses both turning points $z_{1,2}$ and both poles $z_\pm$, where the desired Voros symbol can be recovered by subtracting the residue contributions associated with the enclosed $z_{\pm}$ poles. This integration cycle is depicted in figure~\ref{fig:Integration_Cycle_V12}.\\ 

\noindent 
The relevant residues can be readily evaluated using the general relation~\eqref{eq:Residue_Formula_EWKB}. With the branch-cut assignment dictated by the asymptotic behavior~\eqref{eq:Branch_Assignment}, together with the choice of a branch cut connecting $z_1$ and $z_2$, and hence the pole residues $c_\pm$ given in equation~\eqref{eq:Residues_Horizions}, we obtain
\begin{align}
	\!\!\mathop{\mathrm{Res}}_{\;\;\,z=z_\pm}\!\!\!\!\;\big[S(z,\eta)\big] &= \!\!\mathop{\mathrm{Res}}_{\;\;\,z=z_\pm}\!\!\!\!\;\big[S_\mathrm{even}(z,\eta)\big] + \!\!\mathop{\mathrm{Res}}_{\;\;\,z=z_\pm}\!\!\!\!\;\big[S_\mathrm{odd}(z,\eta)\big] \notag \\ 
	&= \frac{1}{2} + \frac{1}{2} \sqrt{\eta^2\cbig[\Omega^2(1\pm \varepsilon)^4-1\cbig]+1} \, ,
\end{align}
where we simplified the square roots using $\operatorname{Re}(c_\pm)>0$ and $\eta > 0$ and recognized the fact that the logarithmic derivative $S_\mathrm{even}(z,\eta)=-\frac{1}{2}\partial_z \log\!\big[S_\mathrm{odd}(z,\eta)\big]$ has residue $\frac{1}{2}$. The Voros symbol is therefore given by 
\begin{align}
	\mathcal{V}_{1\!\!\;2} &= -\mathcal{S}\exp\cbigg\{\mathlarger{\oint}_{\gamma_{1\!\!\;2}} S(z,\eta)\,\mathrm{d}z \cbigg\} \notag \\ 
	&= -\mathcal{S}\exp\cbigg\{\mathlarger{\oint}_{\gamma_{\mathrm{tot}}} S(z,\eta)\,\mathrm{d}z - 2\pi i \cBig(\!\!\mathop{\mathrm{Res}}_{\;\;\,z=z_+}\!\!\!\!\;\big[S(z,\eta)\big]+\!\!\mathop{\mathrm{Res}}_{\;\;\,z=z_-}\!\!\!\!\;\big[S(z,\eta)\big]\!\!\:\cBig)\!\cbigg\} \\
	&= -\mathcal{S}\exp\cbigg\{\mathlarger{\oint}_{\gamma_{\mathrm{tot}}} S(z,\eta)\,\mathrm{d}z -i\pi \bigg(\!\!\:\sqrt{\eta^2\cbig[\Omega^2(1+\varepsilon)^4-1\cbig]+1}+\sqrt{\eta^2\cbig[\Omega^2(1-\varepsilon)^4-1\cbig]+1}\:\!\bigg)\!\cbigg\} ,  \notag 
\end{align}
where in the first line, we have utilized equation~\eqref{eq:Relation_Voros_S} to relate the cycle over $S_\mathrm{odd}(z,\eta)$ to the cycle over the full $S(z,\eta)$. \\

\noindent
This only leaves the circular contour integral of $S(z,\eta)$ over $\gamma_{\mathrm{tot}}$ to be evaluated. Rather than trying to reconstruct the Voros symbol from the asymptotic series~\eqref{eq:SSeries} of $S(z,\eta)$, expanding the Riccati equation~\eqref{eq:Riccati_Equation} in $\varepsilon$ instead allows us to obtain the all-order result in $\eta$, for any desired order in $\varepsilon$. \\

\noindent 
We therefore return to the Riccati equation~\eqref{eq:Riccati_Equation} defining $S(z,\eta)$ and seek a solution organized in nonnegative powers of $\varepsilon$. The $\varepsilon$-expansion of the WKB potential reads
\begin{align}
	Q(z)&=\frac{j(j+1)}{\big(z-1-\varepsilon\big)\big(z-1+\varepsilon\big)} +\frac{\varepsilon^2\big(\Omega^2 z^4-1\big)}{\big(z-1-\varepsilon\big)^{\!\!\:2}\big(z-1+\varepsilon\big)^{\!\!\:2}} \notag \\[0.05cm] 
	&= \frac{j(j+1)}{(z-1)^2}+ \frac{j(j+1)+\Omega^2 z^4-1}{(z-1)^4}\,\varepsilon^2+ \frac{j(j+1)+2\big(\Omega^2 z^4-1\big)}{(z-1)^6}\,\varepsilon^4 + \mathcal{O}\big(\varepsilon^6\big)\, .
\end{align}
Here, the expansion requires $z-1=\mathcal{O}(1)$ along the integration contour. This condition is ensured by choosing $\gamma_{\mathrm{tot}}$ to be a circle centered at $z_{\mathrm{mid}}=1$ with radius of $\mathcal{O}(1)$. Working to the third nontrivial order, we insert the ansatz 
\begin{align}
	S(z,\eta)=s_0(z,\eta)+s_2(z,\eta)\varepsilon^2+s_4(z,\eta)\varepsilon^4+\mathcal{O}\big(\varepsilon^6\big)
\end{align}
into the Riccati equation~\eqref{eq:Riccati_Equation}. Solving order by order in $\varepsilon$ gives the relations
\begin{subequations}
	\begin{align}
		s_0(z,\eta)^2 + s_0'(z,\eta)&=\eta^2 \,\frac{j(j+1)}{(z-1)^2}\, , \label{eq:ODE_s0} \\ 
		s_2'(z,\eta) + 2s_0(z,\eta) s_2(z,\eta)&= \eta^2\,\frac{j(j+1)+\Omega^2 z^4-1}{(z-1)^4}\, , \label{eq:ODE_s2} \\ 
		s_4'(z,\eta) + 2s_0(z,\eta) s_4(z,\eta)&= \eta^2\,\frac{j(j+1)+2\big(\Omega^2 z^4-1\big)}{(z-1)^6}  - s_2(z,\eta)^2\, .
	\end{align}
    \label{eq:Riccati_Expansion}%
\end{subequations}
Remarkably, these equations can be solved exactly in $\eta$, without resorting to the formal large-$\eta$ expansion characteristic of the standard WKB construction. We can therefore determine each $s_{2n}(z,\eta)$ in a form that retains the full dependence on $\eta$ and thus effectively resums the corresponding WKB series to all orders! \\

\noindent
The resulting solutions contain integration constants that are not fixed by the differential equations~\eqref{eq:Riccati_Expansion} alone. These constants can be determined by expanding the solutions at large $\eta$ and matching the result to the formal WKB expansion. This is illustrated most easily for $s_0(z,\eta)$, as the most general solution of the ODE~\eqref{eq:ODE_s0} takes the form 
\begin{align}
	\!\!\!\:s_0(z,\eta)=\frac{1+\sqrt{1+4\eta^2j(j+1)}}{2(z-1)}- \frac{\sqrt{1+4\eta^2j(j+1)}}{z-1} \cBig[1+c_0(z-1)^{\sqrt{1+4\eta^2j(j+1)}}\cBig]^{\!\!\:-1}\, ,
	\label{eq:General_Solution_s0}
\end{align}
with $c_0\in \mathbb{C}$ being an initially arbitrary integration constant. The role of $c_0$ becomes transparent upon recalling $s_0=\Psi_0'/\Psi_0$. The corresponding solution of the linear Schr\"odinger-type equation is
\begin{align}
	\Psi_0(z,\eta)&= a_0(z-1)^{\frac{1}{2}-\frac{1}{2}\sqrt{1+4\eta^2j(j+1)}} + b_0(z-1)^{\frac{1}{2}+\frac{1}{2}\sqrt{1+4\eta^2j(j+1)}} \notag \\ 
    &\propto (z-1)^{\frac{1}{2}-\frac{1}{2}\sqrt{1+4\eta^2j(j+1)}} \cBig[1+ \frac{b_0}{a_0} \:\!(z-1)^{\sqrt{1+4\eta^2j(j+1)}}\cBig] .
\end{align}
We can thus infer that the constant $c_0$ corresponds to $b_0/a_0$. Since the two terms $\smash{(z-1)^{\frac{1}{2}\pm\frac{1}{2}\sqrt{1+4\eta^2j(j+1)}}}$ precisely constitute the two independent WKB branches, a finite nonzero $c_0$ amounts to mixing these two branches. Such a mixture introduces a dependence in $\Psi_0$ of the form $\smash{(z-1)^{\sqrt{1+4\eta^2j(j+1)}}\sim \exp\cbig[2\eta\sqrt{j(j+1)}\log(z-1)\cbig]}$, which is not captured by the assumed expansion in integer powers of $1/\eta$. Restricting the solution to a single WKB branch therefore requires either $b_0=0$ or $a_0=0$, corresponding to $c_0=0$ or $c_0=\infty$ respectively. With the branch choices established above, we arrive at the all-order WKB result 
\begin{align}
	s_0(z,\eta)=\frac{1+ \sqrt{1+4\eta^2j(j+1)}}{2(z-1)}\, ,
	\label{eq:WKB_Solution_s0}
\end{align}
where the positive square root is selected by requiring $S(z,\eta)>0$ for sufficiently large $z$.\\ 

\noindent 
Inserting the solution~\eqref{eq:WKB_Solution_s0} into relation~\eqref{eq:ODE_s2} yields the general second-order solution as 
\begin{align}
	s_2(z,\eta)= \eta^2\Bigg\{&\frac{\Omega^2 +j(j+1)-1}{\sqrt{1+4\eta^2 j(j+1)}-2}\,(z-1)^{-3} +\frac{4\Omega^2}{\sqrt{1+4\eta^2 j(j+1)}-1} \,(z-1)^{-2} \notag \\ 
    +\,&\frac{6\Omega^2}{\sqrt{1+4\eta^2 j(j+1)}-0} \,(z-1)^{-1} +\frac{4\Omega^2}{\sqrt{1+4\eta^2j(j+1)}+1}\,(z-1)^0 \notag \\ 
	+\,&\frac{\Omega^2}{\sqrt{1+4\eta^2j(j+1)}+2}\,(z-1)^1\Bigg\} + c_2 (z-1)^{-1-\sqrt{1+4\eta^2j(j+1)}}\, .
\end{align}
%We have deliberately left certain terms unsimplified in order to make the structure of the solution more transparent. 
As before, consistency with the chosen WKB branch requires the homogeneous contribution proportional to $\smash{(z-1)^{-1-\sqrt{1+4\eta^2j(j+1)}}}$ to be absent, thereby fixing $c_2=0$. Recall that the case $j=0$ has been excluded from our analysis, and thus the pole arising from $\sqrt{1+4\eta^2j(j+1)} - 1 \xrightarrow{j=0} 0$ does not pose a problem here. Other positive integer values of $j$ do not give rise to additional difficulties. \\ 

\noindent 
From the discussion of the first two nontrivial orders, it becomes evident that the desired WKB solution $S(z,\eta)$ takes the form of a Laurent series in $z-1$, with
\begin{align}
	s_{2k}(z,\eta) = \frac{1}{z-1}\!\!\:\sum_{\ell=-2k}^{2k} \!\alpha_{\ell}^{(2k)} (z-1)^{\ell} \, .
	\label{eq:Ansatz_s2n}
\end{align}
Care must be taken if $\sqrt{1+4\eta^2j(j+1)}$ assumes integer values, as already encountered in the discussion of $s_2(z,\eta)$. At these isolated values of $\eta$, the Riccati problem becomes resonant and logarithmic contributions involving $\log(z-1)$ appear. However, such terms are absent for generic $\eta$ in a punctured neighborhood of $\eta=1$, where the Riccati coefficients remain meromorphic in $z$ and the sought-after contour integral over $\gamma_{\mathrm{tot}}$ is unambiguously defined. We therefore evaluate the relevant contour integral for generic, nonresonant $\eta$ and take the limit $\eta\to 1$ only after obtaining the desired Voros symbol. Assuming the resulting expression admits a finite and smooth limit, the isolated resonant values encountered at intermediate stages do not affect this continuation and need not be treated separately. This prescription is supported by validating the arising ZDM spectrum with direct numerical computations, see section~\ref{sec:Numerical_Crosschecks}. In cases where the limiting expression instead develops a singularity, the resonant problem must be treated explicitly, including the associated logarithmic contributions.\\

\noindent 
Restricting attention to the nonresonant case, in which the desired Riccati solution takes the form~\eqref{eq:Ansatz_s2n}, the contour integral over $\gamma_\mathrm{tot}$ is determined entirely by the residue at $z=1$, \textit{i.e.} by the coefficient of $(z-1)^{-1}$ in $S(z,\eta)$. This coefficient can be obtained directly from the Riccati equation by inserting the ansatz~\eqref{eq:Ansatz_s2n} and solving the resulting algebraic system for the coefficients $\alpha_{\ell}^{(2k)}$ order by order in $z-1$. Stating only the relevant coefficient of $s_4(z,\eta)$, we arrive at
\begin{align}
	\alpha_0^{(4)}&=\big[(z-1)^{-1}\big]s_4(z,\eta) \notag \\ 
	&= \frac{2\eta^2\Omega^2}{\sqrt{1+4\eta^2j(j+1)}} \cBigg\{1-\eta^2\cbigg[\frac{18\Omega^2}{4\eta^2j(j+1)+1}+\frac{16\Omega^2}{4\eta^2j(j+1)} \notag \\[-0.1cm] 
	&\qquad\qquad\qquad\qquad\qquad\qquad\qquad\qquad\qquad\qquad\: +\frac{\Omega^2+j(j+1)-1}{4\eta^2j(j+1)-3}\cbigg]\!\!\:\cBigg\}\, .
\end{align}
The only pole that persists in the final result occurs for $j=0$, which has already been excluded from our analysis. Although the full solution $s_4(z,\eta)$ contains an additional pole associated with resonant enhancements for $j=1$ as $\eta\to 1$, this pole does not contribute to the final result and is therefore immaterial within the continuation prescription described above.\footnote{
For any fixed non-negative integer $j$, $s_{2j+2}(z,\eta)$ is the first function in the near-extremal expansion to develop a resonant pole. However, as is already apparent for $j=0$, this pole does not initially occur in the coefficient of $(z-1)^{-1}$ and therefore does not affect the contour integral. The resonant pole first propagates into the $(z-1)^{-1}$ coefficient at order $s_{4j+4}(z,\eta)$, at which point the continuation prescription described above must be modified to account explicitly for the resonant solution. Consequently, for fixed $j$, the Voros symbol $\mathcal{V}_{12}$ can be obtained by the present procedure without encountering such a singularity through order $\varepsilon^{4j+2}$. The first problematic contribution therefore occurs at order $\varepsilon^{4j+4}$.} \\ 

\noindent 
Truncating the expansion at this order, as higher-order contributions become increasingly cumbersome without providing additional insight, we obtain the sought-after contour integral as
\begin{align}
	\frac{1}{2\pi i}\mathlarger{\oint}_{\gamma_{\mathrm{tot}}} S\big(z,\eta\!=\!1\big)\,\mathrm{d}z &= j+1 + \frac{6\Omega^2\varepsilon^2}{2j+1} + \frac{2\Omega^2 \varepsilon^4}{2j+1} \Bigg\{1-\frac{18\Omega^2}{4j(j+1)+1} \label{eq:Integral_Voros_Symb_12}\\ 
	&\qquad\qquad\qquad\qquad\!\!\!\: -\frac{16\Omega^2}{4j(j+1)} -\frac{\Omega^2+j(j+1)-1}{4j(j+1)-3}\Bigg\} +\:\! \mathcal{O}\big(\varepsilon^6\big)\, . \notag
\end{align}
Here, we have finally set $\eta=1$, which reduces the square-root factor according to $\smash{\sqrt{1+4\eta^2 j(j+1)}\xrightarrow{\eta \to 1} 2j+1}$. Likewise, upon setting $\eta=1$, the Voros symbol $\mathcal{V}_{1\!\!\;2}$ takes the form
\begin{align}
		\big[\mathcal{V}_{1\!\!\;2}\big]_{\eta=1} &= - \exp\cbigg\{\mathlarger{\oint}_{\gamma_{\mathrm{tot}}} S\big(z,\eta\!=\!1\big)\,\mathrm{d}z -2i\pi \Omega \big(1+\varepsilon^2\big) \!\cbigg\}\, , 
		\label{eq:Voros_Symbol_12}
\end{align}
where $\Omega>0$ allows us to choose the branch $\sqrt{\Omega^2}=\Omega$. Recall that the case $\Omega<0$ has been excluded from the analysis, as it never satisfies the prescribed QNM boundary conditions.

\subsection{Estimate of $\mathcal{V}_{23}$}
\label{sec:Voros_Symbol_23}

A precise evaluation of $\mathcal{V}_{23}$ is considerably more involved than that of $\mathcal{V}_{12}$, because the two turning points defining the cycle are parametrically separated as
\begin{align}
	z_2-1&=\mathcal{O}(\varepsilon)\, , & z_3&=\mathcal{O}\big(\varepsilon^{-1}\big)\, .
\end{align}
A proper determination of the Voros symbol $\mathcal{V}_{23}$ would therefore require a matched asymptotic analysis involving inner, intermediate, and outer radial regions:
\begin{subequations}
	\begin{align}
		\text{inner regime: }& \big\lvert z-1\big\rvert = \mathcal{O}(\varepsilon)\, , \\ 
		\text{intermediate regime: }& \varepsilon \ll \big\lvert z-1\big\rvert \ll \varepsilon^{-1}\, , \\
		\text{outer regime: }& \big\lvert z-1\big\rvert =\mathcal{O}\big(\varepsilon^{-1}\big) \, .
	\end{align}
\end{subequations}
A full matching procedure is unnecessary for our purposes, however, since we only require the leading-order dependence on $\varepsilon$. As we have previously established in equation~\eqref{eq:WKB_Solution_s0}, in the parametrically broad intermediate region $\varepsilon \ll \big\lvert z-1\big\rvert \ll \varepsilon^{-1}$, the two relevant Riccati branches reduce, at leading order in $\varepsilon$, to
\begin{align}
	S^{(\pm)}(z,\eta) = \frac{1\pm \sqrt{1+4\eta^2j(j+1)}}{2(z-1)} + \text{subleading}\, .
	\label{eq:LO_Riccati_Intermediate}
\end{align}
At the level of logarithmic accuracy, the intermediate contribution may therefore be estimated directly from the parametric locations of the two turning points as
\begin{align}
	\mathlarger{\oint}_{\gamma_{2\!\!\;3}} S(z,\eta)\,\mathrm{d}z &= \mathlarger{\int}_{z_2}^{z_3} \Big[S^{(+)}(z,\eta)-S^{(-)}(z,\eta)\Big] \,\mathrm{d}z \notag \\ 
	&= \sqrt{1+4\eta^2j(j+1)} \mathlarger{\int}_{\scalebox{0.7}{$1\!\!\:+\!\!\:\mathcal{O}(\varepsilon)$}}^{\scalebox{0.7}{$\,\mathcal{O}\big(\varepsilon^{-1}\big)$}} \,\frac{\mathrm{d}z}{z-1} + \mathcal{O}(1) \notag \\
	&= 2\sqrt{1+4\eta^2j(j+1)} \,\log\!\big(\varepsilon^{-1}\big) + \mathcal{O}(1)\, .
	\label{eq:LeadingScaling_V23_Exponent}
\end{align}
Crucially, the inner and outer regions contribute only to the finite $\mathcal{O}(1)$ part of the Voros exponent, which we leave implicit. After the respective rescalings $z-1=\varepsilon y$ and $z-1=\tilde{y}/\varepsilon$, both the integration domains and the corresponding rescaled integrands are of order unity. Neither region therefore generates a logarithmic enhancement as $\varepsilon\to 0$. Determining this finite contribution would require carrying out the full matching procedure, which is unnecessary for the present purpose.\footnote{A systematic treatment would introduce matching points $z_\mathrm{inner}$ and $z_\mathrm{outer}$, defined by $\lvert z_\mathrm{inner}-1\rvert =A\varepsilon$ and $\lvert z_\mathrm{outer}-1\rvert=B\varepsilon^{-1}$, with $A\gg 1$ and $B\ll 1$. The integral would then be evaluated separately in the three scaling regions and the resulting asymptotic expansions matched in their respective overlap domains, with the auxiliary dependence on $A$ and $B$ canceling in the final result. For a systematic discussion of the matching procedure and its underlying asymptotic arguments, we refer the reader to the classic textbook by Bender and Orszag~\cite{Bender1999advanced}.} Exponentiating the previous result~\eqref{eq:LeadingScaling_V23_Exponent}, the Voros symbol is found to scale as \begin{align}
	\mathcal{V}_{23}&= \varepsilon^{-2\sqrt{1+4\eta^2j(j+1)}} \times \mathcal{O}(1) \xrightarrow{\eta\to 1} \varepsilon^{-2(2j+1)} \times \mathcal{O}(1) \, ,
	\label{eq:LeadingScaling_V23}
\end{align}
and is therefore parametrically enhanced in the near-extremal limit $\varepsilon\to 0$. This scaling considerably simplifies the quantization condition, as we will discuss in the following subsection.

\subsection{Solving for the ZDM spectrum}

Before turning to the EQC~\eqref{eq:EQC_RN_Total} for $\Omega^2>j(j+1)+1$, which will ultimately yield the desired ZDM spectrum, let us first show that the alternative EQC~\eqref{eq:EQC_RN_Total_2} admits no solutions within the regime of validity of our analysis. To this end, note that condition~\eqref{eq:EQC_RN_Total_2} requires $\mathcal{V}_{23}=-\mathcal{R}^{-1}$. For real $\Omega$, however, expression~\eqref{eq:Ratio_R} implies $\lvert\mathcal{R}\rvert=\mathcal{O}(1)$, whereas equation~\eqref{eq:LeadingScaling_V23} gives $\lvert\mathcal{V}_{23}\rvert=\mathcal{O}\big(\varepsilon^{-4j-2}\big)$. Reconciling these incompatible scalings would require $\mathcal{R}^{-1}$ to acquire an equally strong parametric enhancement in the near-extremal limit $\varepsilon\to 0$, which could only arise by allowing $\Omega$ to develop a logarithmically growing imaginary part, signifying behavior outside the assumptions under which the EQC was derived and is therefore inconsistent. We thus conclude that the branch $\Omega^2<j(j+1)+1$ yields no admissible ZDM solutions and can be discarded. \\

\noindent 
This lets us turn our attention to the physical case $\Omega^2>j(j+1)+1$, for which the EQC~\eqref{eq:EQC_RN_Total} can be recast into the convenient form 
\begin{align}
	1+\mathcal{V}_{12}=-\big(\mathcal{R}\mathcal{V}_{23}\big)^{-1}=\mathcal{O}\big(\varepsilon^{4j+2}\big)\, .
    \label{eq:EQCfinalform}
\end{align}
Since $\mathcal{V}_{23}$ is parametrically enhanced in the near-extremal limit, its inverse is correspondingly strongly suppressed. This also explains retrospectively why the $\mathcal{O}(1)$ prefactor of $\mathcal{V}_{23}$ need not be fixed for our present purpose, as it can affect the quantization condition only at order $\varepsilon^{4j+2}$. In other words, when solving for the QNM spectrum below this order, we can self-consistently neglect the right-hand side of equation~\eqref{eq:EQCfinalform}, which reduces to $1+\mathcal{V}_{12}=0$ within the required accuracy.\footnote{Interestingly, an independent limitation arises from the expansion of $\mathcal{V}_{12}$ itself. As briefly anticipated above, the na\"{\i}ve nonresonant construction employed in section~\ref{sec:Voros_Symbol_12} remains sufficient only up to a finite order for fixed integer $j$. In particular, $s_{4j+4}$ is the first term for which a resonant pole survives in the coefficient of $(z-1)^{-1}$, and hence in the contour integral determining $\mathcal{V}_{12}$. From this order onward, the resonant Riccati problem must be treated explicitly, including the associated logarithmic contributions. The presented construction therefore determines $\mathcal{V}_{12}$ straightforwardly through order $\varepsilon^{4j+2}$, with the first problematic contribution arising at order $\varepsilon^{4j+4}$. Remarkably, this intrinsic limitation coincides with the accuracy beyond which the previously neglected contribution from $\big(\mathcal{V}_{23}\big)^{-1}$ must also be taken into account. Thus, both sides of the EQC become sensitive to additional structure at essentially the same order.} \\

\noindent 
We will thus restrict our analysis to the first few orders in the near-extremal expansion. Given $j \geq 1$, all terms obtained in equation~\eqref{eq:Integral_Voros_Symb_12} through $\mathcal{O} \big( \varepsilon^4 \big)$ lie below the threshold $\mathcal{O} \big( \varepsilon^{4j+2} \big)$, consistent with neglecting the $\mathcal{O}(1)$ prefactor of $\mathcal{V}_{23}$. The reduced quantization condition $\mathcal{V}_{12} = -1$ fixes the phase of $\mathcal{V}_{12}$ to an odd multiple of $\pi$, which we parametrize as $i\pi(1-2n)$ with $n\in \mathbb{Z}$, and thus equation~\eqref{eq:Voros_Symbol_12} yields
\begin{align}
	\Omega_n \big(1+ \varepsilon^2\big)=n+\frac{1}{2\pi i} \mathlarger{\oint}_{\gamma_{\mathrm{tot}}} S\big(z,\eta\!=\!1\big)\,\mathrm{d}z \, , \qquad n\in \mathbb{Z}\, .
    \label{eq:QuantizationCondition_Intermediate}
\end{align}
Inserting the previous result~\eqref{eq:Integral_Voros_Symb_12} and recalling the accuracy estimate arising from the neglected $\mathcal{V}_{23}$, which enters no earlier than $\mathcal{O}\big(\varepsilon^6\big)$ for $j\geq 1$, we obtain
\begin{align}
	\Omega_n \big(1+ \varepsilon^2\big)=\big(n+j+1\big) +\frac{6\Omega_n^2 \varepsilon^2}{2j+1}+\frac{2\Omega_n^2\varepsilon^4}{2j+1}\Bigg\{1&-\frac{18\Omega_n^2}{4j(j+1)+1} -\frac{16\Omega_n^2}{4j(j+1)} \nonumber \\ 
    &-\frac{\Omega_n^2+j(j+1)-1}{4j(j+1)-3}\Bigg\} + \mathcal{O}\big(\varepsilon^6\big)\, .
\end{align}
This implicit relation can now be solved order by order in $\varepsilon$. At leading order, we immediately recover $\Omega_n=n+j+1+\mathcal{O}\big(\varepsilon^2\big)$, in agreement with the expected near-extremal ZDM spectrum~\eqref{eq:Expected_ZDM_Spectrum}~\cite{Kim:2001ev,Hod:2010hw, Chen:2012zn, Kim:2012mh, Eniceicu:2019npi}. Going beyond the leading order, we obtain 
\begin{align}
	\Omega_n=\big(n+j+1\big)\cBigg\{1 &- \cbigg[1-\frac{6(n+j+1)}{2j+1}\cbigg]\:\!\varepsilon^2 \notag \\ 
	&+\cbigg[1-\frac{n+j+1}{2(2j+1)}\bigg(33-\frac{1}{4j(j+1)-3}\bigg)+\frac{72(n+j+1)^2}{(2j+1)^2} 
    \label{eq:ZDM_Result_Omega_Full} \\ 
	&\qquad\; -\frac{2(n+j+1)^3}{(2j+1)^3}\bigg(35+\frac{4}{j(j+1)}+\frac{4}{4j(j+1)-3}\bigg)\cbigg]\varepsilon^4 + \mathcal{O}\big(\varepsilon^6\big)\cBigg\}. \notag 
\end{align}
Defining the prominent parametric combinations $N=n+j+1$ and $J=2j+1$ and recalling our original rescaling $\Omega = iM\varepsilon^{-1}\omega$, we can express our central result in the compressed form 
\begin{align}
	\omega^{(\mathrm{ZDM})}_n=-\frac{i\varepsilon N}{M}\cBigg\{1 &- \bigg(1-\frac{6N}{J}\bigg)\:\!\varepsilon^2 +\cbigg[1-\frac{N}{2J}\bigg(33-\frac{1}{J^2-4}\bigg)+\frac{72N^2}{J^2} \label{eq:ZDM_Spectrum_NNLO} \\[-0.1cm] 
	&\qquad\qquad\qquad\qquad\quad\;\:\! -\frac{2N^3}{J^3}\bigg(35+\frac{16}{J^2-1}+\frac{4}{J^2-4}\bigg)\cbigg]\varepsilon^4 + \mathcal{O}\big(\varepsilon^6\big)\cBigg\}\, . \notag 
\end{align}
As stated at the outset, we require $\Omega$ (and hence $N$) to remain of order unity, {\it i.e.} to carry no explicit $\varepsilon$-scaling. Otherwise, the power counting employed throughout the analysis would cease to be consistent. This restriction is also manifest in the spectrum above: although formally organized as an expansion in $\varepsilon$, the actual expansion parameter is effectively $\varepsilon N$, since successive orders in $\varepsilon$ are accompanied by increasingly high powers of $N$. While these corrections are suppressed at large angular momentum $j$, sufficiently highly excited modes, corresponding to large $n$, eventually invalidate the perturbative expansion.\\ 

\noindent 
As we have seen, the algebraic solutions~\eqref{eq:ZDM_Spectrum_NNLO} to the EQC~\eqref{eq:EQC_RN_Total} formally admit $n\in \mathbb{Z}$. However, just as in the harmonic oscillator case, not all of these solutions belong to the physical ZDM spectrum. As shown in section~\ref{sec:Case2}, $\Omega$ must satisfy $\Omega^2>j(j+1)+1$, whereas the EQC~\eqref{eq:EQC_RN_Total_2} arising for $\Omega^2<j(j+1)+1$ has no admissible ZDM solutions. This already restricts $n$ to the ranges $n\geq 0$ and $n\leq -2(j+1)$. Additionally, $\Omega$ is required to be positive, as the demanded ZDM boundary conditions~\eqref{eq:QNM_BoundaryConditions_RN} cannot be simultaneously satisfied for $\Omega<0$, as we showed in section~\ref{sec:StokesEQC}. This indeed gives the usual restriction of the spectrum to non-negative integers $n\in\mathbb{N}_0$~\cite{Kim:2001ev,Hod:2010hw, Chen:2012zn, Kim:2012mh, Eniceicu:2019npi}. Note, however, that the result~\eqref{eq:ZDM_Spectrum_NNLO} applies only to angular momentum sectors with $j\geq 1$, as the $s$-wave channel must be treated separately and will be left for future work. As an aside, note that setting $j=0$ in the leading-order term of equation~\eqref{eq:ZDM_Result_Omega_Full} yields the standard result, even though the derivation leading to it is not valid for $j=0$. This agreement should therefore be regarded as accidental rather than as a consistent prediction of the present analysis.

\begin{figure}[h]
    \centering
    \begin{minipage}{0.49\textwidth}
        \includegraphics[width=\linewidth]{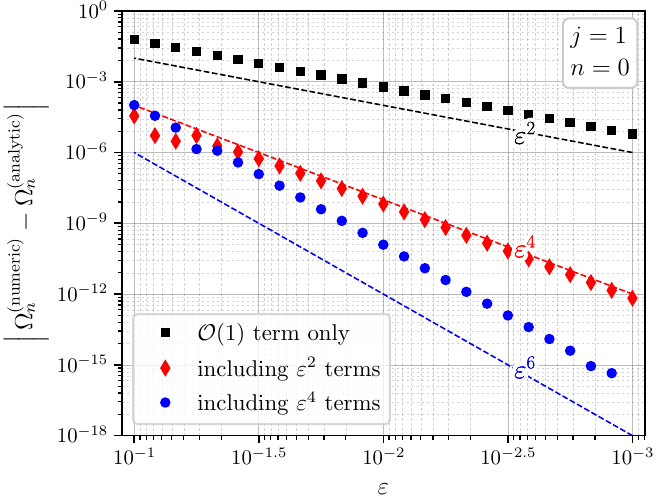}
    \end{minipage}
    \hfill
    \begin{minipage}{0.49\textwidth}
        \includegraphics[width=\linewidth]{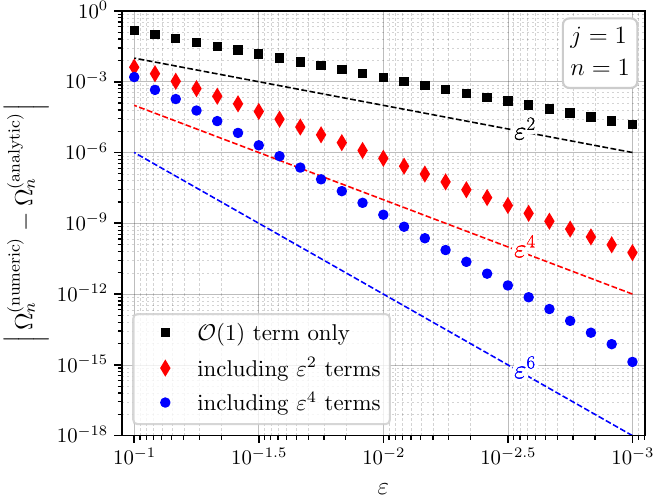}
    \end{minipage}
    \begin{minipage}{0.49\textwidth}
        \includegraphics[width=\linewidth]{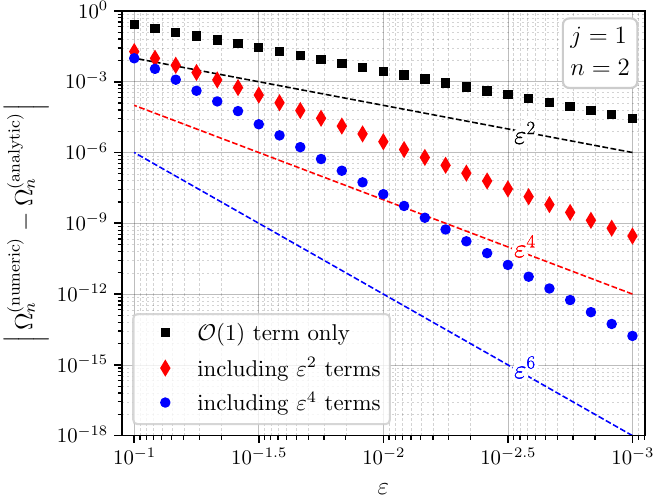}
    \end{minipage}
    \hfill
    \begin{minipage}{0.49\textwidth}
        \includegraphics[width=\linewidth]{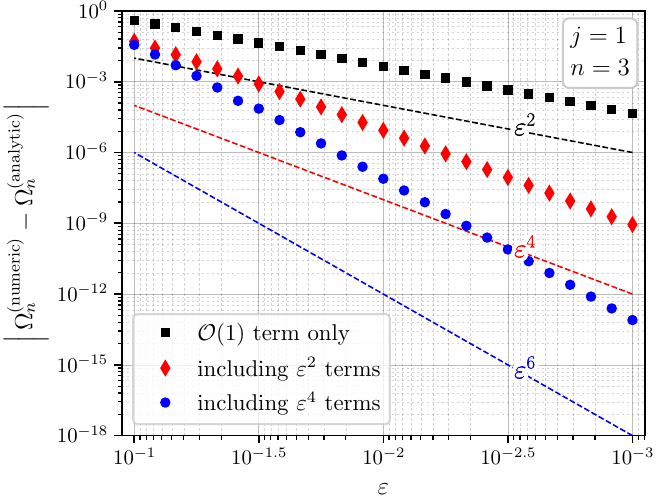}
    \end{minipage}
    \caption{Absolute deviation between the numerically evaluated ZDM frequencies 
    and the analytic expression presented in equation~\eqref{eq:ZDM_Result_Omega_Full}, truncated at $\mathcal{O}(1)$ (black squares), $\mathcal{O}(\varepsilon^2)$ (red diamonds), and $\mathcal{O}(\varepsilon^4)$ (blue circles). All panels refer to the same value $j=1$, but to different overtones $n = 0,1,2,3$.  
    The expected asymptotic behavior in the small-$\varepsilon$ regime, indicated by the dashed lines, is clearly visible, as the scattered data points become parallel to the reference dashed lines corresponding to pure $\varepsilon^2,\varepsilon^4$, and $\varepsilon^6$ scalings, respectively. As anticipated, the overall error increases with $n$.}
    \label{fig:ZDM_Comparison_j0}
\end{figure}

\subsection{Numerical cross-checks}
\label{sec:Numerical_Crosschecks}

\begin{figure}[h]
    \centering
    \begin{minipage}{0.49\textwidth}
        \includegraphics[width=\linewidth]{NumericalComparison_j_1_n_2}
    \end{minipage}
    \hfill
    \begin{minipage}{0.49\textwidth}
        \includegraphics[width=\linewidth]{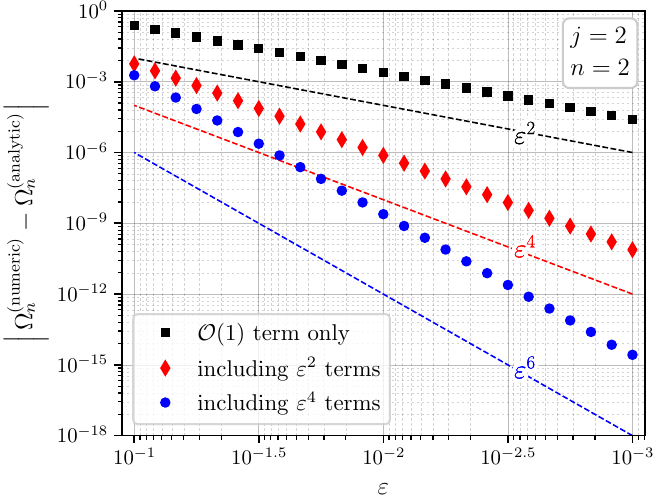}
    \end{minipage}
    \begin{minipage}{0.49\textwidth}
        \includegraphics[width=\linewidth]{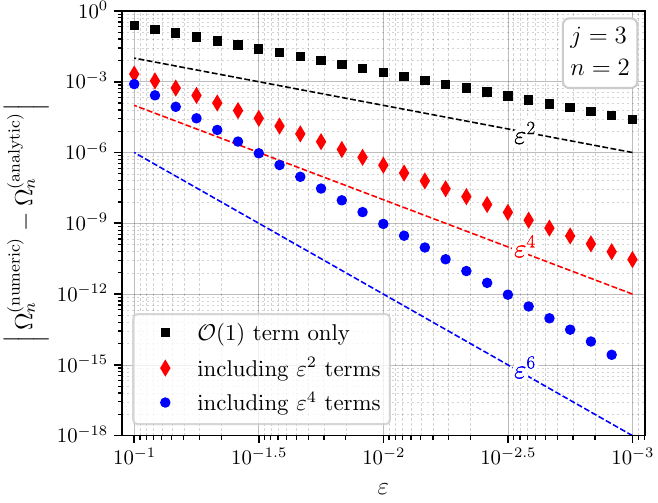}
    \end{minipage}
    \hfill
    \begin{minipage}{0.49\textwidth}
        \includegraphics[width=\linewidth]{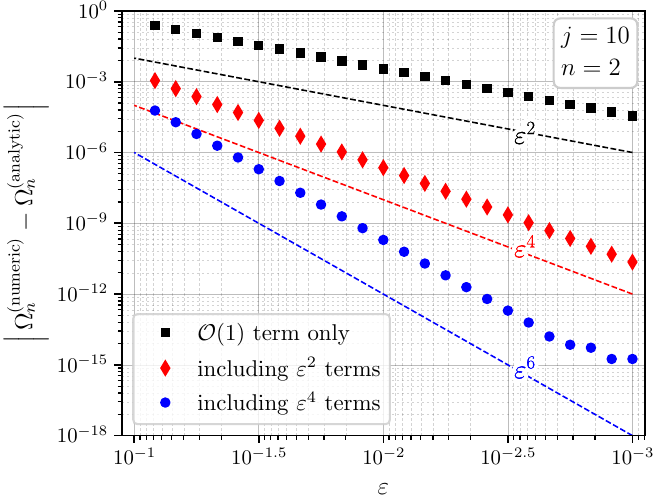}
    \end{minipage}
    \caption{ Same as figure~\ref{fig:ZDM_Comparison_j0}, but for fixed $n=2$ and varying $j=1,2,3,10$. Again, the numerical results consistently validate the analytic prediction. At very small and very large $\varepsilon$, numerical instabilities become dominant and lead to apparent departures from the expected linear scaling, most notably in the case of large $j=10$. Missing data points correspond to parameter values for which the numerical solver failed to converge at all.}
    \label{fig:ZDM_Comparison_n2}
\end{figure}

Before concluding, let us verify that for $j\neq 0$, our central result~\eqref{eq:ZDM_Spectrum_NNLO} not only reproduces the previously known leading-order expression~\eqref{eq:Expected_Result} for the ZDM frequencies $\omega_n^{(\mathrm{ZDM})}$, but also agrees at the next two orders with the results obtained from a purely numerical analysis. To this end, we employ Leaver's continued-fraction method~\cite{Leaver:1985ax, Leaver:1990zz, Berti:2009kk, Konoplya:2011qq}, which provides a highly accurate numerical determination of the ZDM spectrum, with accuracy limited only by numerical precision. The comparison is shown in figures~\ref{fig:ZDM_Comparison_j0} and \ref{fig:ZDM_Comparison_n2}, where we show the absolute deviation of the analytic result from the numerical results for different orders in $\varepsilon$ and different values of $n$ and $j$. 
As is readily seen, the numerical results confirm the analytic expression~\eqref{eq:ZDM_Result_Omega_Full}, with each successive order exhibiting the expected asymptotic scaling as $\varepsilon$ decreases.

\newpage 
\section{Conclusions}
\label{sec:Outlook}

In this paper, we have computed the ZDM spectrum of a massless scalar field propagating in a near-extremal RN background using exact WKB methods. Starting from the Stokes geometry shown in figure~\ref{fig:StokesGeometry_Relevant}, we have determined the associated EQC~\eqref{eq:EQC_RN_Total}. Evaluating this EQC requires determining the Voros symbols encountered in the problem, which are computed in section~\ref{sec:Spectrum_Computation}
via a suitable deformation of the relevant integration contours and a treatment that manifestly retains all orders in the WKB expansion parameter $\eta$. This leaves us with only an expansion in the near-extremality parameter $\varepsilon$, in contrast to previous exact WKB analyses, which either stop at the first order in $\eta$~\cite{Miyachi:2025ptm} or approximate the all-order results via Padé approximants~\cite{Hatsuda:2026ghx}. Our result, given in equation~\eqref{eq:ZDM_Spectrum_NNLO}, systematically improves on the accuracy of previous analytic calculations~\cite{Kim:2001ev,Hod:2010hw, Chen:2012zn, Kim:2012mh}.
As shown in section~\ref{sec:Spectrum_Computation}, we find agreement with a numerical analysis, with the residuals exhibiting precisely the expected $\varepsilon$-scaling.\\

\noindent 
We note that our treatment still leaves open two unanswered questions, namely how to address the $j=0$ $s$-wave perturbations and how to systematically extend the result to higher orders beyond $\mathcal{O}\big(\varepsilon^{4j}\big)$. We remark that accessing very high orders in the $\varepsilon$-expansion would be a prerequisite for understanding possible resurgent connections with the ``vanilla'' branch of the QNM spectrum.\\

\noindent 
Our method opens a new avenue for understanding QNM spectra of RN BHs and other scenarios of interest. 
In particular, natural directions to explore include massive or charged scalar fields and higher-spin representations in the RN background. Another possibility would be to study the most general BH metric compatible with SM matter, the Kerr--Newman metric. Moreover, this method can also help in understanding the effect of a dark matter halo surrounding the BH~\cite{Barausse:2014tra,Cardoso:2021wlq} and in studies of more general metrics carrying exotic charges~\cite{Achucarro:1995nu,Ayon-Beato:1998hmi,Ayon-Beato:2000mjt,Herdeiro:2014goa,Herdeiro:2015tia,Maldacena:2020skw,Gervalle:2024yxj}.\\

\noindent
Other future work would include a study of the frequency dependence of the EQC and its relation to scattering data, {\it i.e.} the transmission and reflection coefficients of the horizon. This would allow the computation of greybody factors of BHs~\cite{Oshita:2023cjz,Konoplya:2024lir} using exact WKB. Another idea would be to assess whether a possible connection between the weak gravity conjecture and ZDMs~\cite{Hod:2017uqc,Urbano:2018kax,Harlow:2022ich} holds, which necessitates ZDM computations away from extremality~\cite{Harlow:2022ich}. \\

\noindent
In conclusion, we have shown that exact WKB methods can be successfully applied to the study of ZDMs of RN BHs, improving the state of the art in analytical computations.
As outlined above, we hope this work serves as a basis for further analytic studies of phenomenological properties of BHs in many possible regimes.

\section*{Acknowledgments}
NW gratefully acknowledges support from the German Academic Scholarship Foundation, the Marianne--Plehn--Program of the Elite Network of Bavaria, and the International Max Planck Research School on Elementary Particle Physics (IMPRS EPP).  The research of PLC is supported by the Max Planck Society--Weizmann Institute of Science joint postdoctoral program. 
FY is supported by the Cluster of Excellence {\em Precision Physics, Fundamental Interactions and Structure of Matter\/} (PRISMA${}^{++}$ -- EXC~2118/2) within the German Excellence Strategy (project ID 390831469).
The authors thank D. Eniceicu and M. Reece for initial discussions on their paper~\cite{Eniceicu:2019npi}. We furthermore acknowledge the use of LLMs, namely ChatGPT-5.6 Sol and Claude Sonnet 5, for backend coding assistance, cross-checking analytical and numerical computations, manuscript copyediting, and useful discussions.

\addcontentsline{toc}{section}
{\protect\numberline{}References}
\bibliographystyle{JHEP}
\bibliography{Literature.bib}

\end{document}